\documentclass{emulateapj}
\usepackage{natbib}
\usepackage{epsfig, color, subfigure}
\usepackage{amssymb,amsmath, epstopdf}
\usepackage{multirow}
\usepackage{longtable}
\usepackage{hyperref}

\shorttitle{Metallicity profile of M31}
\shortauthors{Sanders et al.}

\newcommand{\CountPNNGC}{23}
\newcommand{\CountHiiTotal}{253}
\newcommand{\CountHiiADC}{200}
\newcommand{\CountPNTotal}{407}
\newcommand{\CountPNADC}{392}
\newcommand{\CountPNhalo}{81}
\newcommand{\CountPNdisk}{326}
\newcommand{\CountNoClass}{25}

\newcommand{\SBmeanRatio}{2.4}
\newcommand{\SBmeanRatioC}{4.4}
\newcommand{\AzimluCompScat}{0.3}
\newcommand{\AzimluFPNa}{2.6}
\newcommand{\AzimluFPNb}{5.4}
\newcommand{\AzimluFHIIa}{7.0}
\newcommand{\AzimluFHIIb}{31.1}

\newcommand{\diffusepne}{7}
\newcommand{\diffusepneper}{1}
\newcommand{\HIINdirect}{4}
\newcommand{\PNdirectmedLone}{8.26}
\newcommand{\PNdirectmedLtwo}{8.42}
\newcommand{\PNdirectmedLthree}{8.49}
\newcommand{\HIImedPTohfive}{8.34\pm0.12}
\newcommand{\HIImedNohsixNtwo}{8.89\pm0.24}

\newcommand{\HIINAv}{199}

\newcommand{\HIIGradientZninefourBboot}{$-0.0208 \pm 0.0048$}

\newcommand{\HIISpearmanZninefourp}{2\times10^{-06}}

\newcommand{\HIIGradientKDohtwoBboot}{$-0.0096 \pm 0.0049$}

\newcommand{\HIISpearmanKDohtwop}{1\times10^{-04}}

\newcommand{\HIIGradientNohsixNtwoBboot}{$-0.0195 \pm 0.0055$}
\newcommand{\HIINNohsixNtwo}{192}

\newcommand{\HIISpearmanNohsixNtwop}{3\times10^{-04}}

\newcommand{\HIIGradientNohsixOthreeNtwoAboot}{$8.98 \pm 0.08$}
\newcommand{\HIIGradientNohsixOthreeNtwoBboot}{$-0.0130 \pm 0.0068$}
\newcommand{\HIINNohsixOthreeNtwo}{100}

\newcommand{\HIISpearmanNohsixOthreeNtwop}{0.06}

\newcommand{\HIIGradientPTohfiveBboot}{$-0.0054 \pm 0.0064$}

\newcommand{\HIISpearmanPTohfivep}{0.60}

\newcommand{\HIIGradientPVTNHONSBboot}{$-0.0303 \pm 0.0049$}
\newcommand{\HIINPVTNHONS}{52}

\newcommand{\HIIstellarGradientNtwoBboot}{$-0.0080 \pm 0.0056$}

\newcommand{\HIIstellarSpearmanZninefourp}{0.06}

\newcommand{\HIIstellarGradientNohsixNtwoAboot}{$9.24 \pm 0.11$}
\newcommand{\HIIstellarGradientNohsixNtwoBboot}{$-0.0224 \pm 0.0082$}

\newcommand{\HIIstellarNPVTNHONS}{10}

\newcommand{\HIIdiffuseGradientNtwoBboot}{$-0.0177 \pm 0.0049$}

\newcommand{\HIIdiffuseSpearmanZninefourp}{6\times10^{-05}}

\newcommand{\HIIdiffuseGradientNohsixNtwoAboot}{$9.09 \pm 0.09$}
\newcommand{\HIIdiffuseGradientNohsixNtwoBboot}{$-0.0195 \pm 0.0070$}

\newcommand{\HIIdiffuseGradientPVTNHONSAboot}{$7.83 \pm 0.08$}
\newcommand{\HIIdiffuseGradientPVTNHONSBboot}{$-0.0300 \pm 0.0058$}

\newcommand{\PNNAv}{333}

\newcommand{\PNNNtwo}{277}

\newcommand{\PNDimBbootNtwo}{$0.0178 \pm 0.0108$}

\newcommand{\PNBrightBbootNtwo}{$-0.0122 \pm 0.0059$}

\newcommand{\PNGradientdirectBboot}{$-0.0056 \pm 0.0076$}
\newcommand{\PNNdirect}{51}

\newcommand{\PNBrightBbootdirect}{$-0.0023 \pm 0.0097$}

\newcommand{\PNSpearmandirectp}{0.45}

\newcommand{\HIINohsixNtwomedian}{8.89}

\newcommand{\HIIPTohfivemedian}{8.34}

\newcommand{\HIINohsixerrNinety}{0.10}
\newcommand{\HIINohsixerrMedian}{0.03}
\newcommand{\PNescatter}{0.26}
\newcommand{\PNemedian}{8.46}
\newcommand{\PNescatterN}{51}
\newcommand{\PNemedianerr}{0.10}

\newcommand{\PNemedianAvN}{333}
\newcommand{\PNeAvFull}{0.47^{+0.76}_{-0.46}}
\newcommand{\PNehaloscatter}{0.22}
\newcommand{\PNehalomedian}{8.50}
\newcommand{\PNehaloscatterN}{17}

\newcommand{\PNehalomedianAvN}{75}
\newcommand{\PNehaloAvFull}{0.08^{+0.19}_{-0.08}}
\newcommand{\PNlocalOHN}{20}

\newcommand{\PNlocalOHtopthirdf}{0.2}

\newcommand{\PNlocalAvN}{502}

\newcommand{\HIIlocalOHN}{132}
\newcommand{\HIIlocalOHdiff}{0.6}
\newcommand{\HIIlocalOHtopthirdf}{0.3}
\newcommand{\HIIlocalAvdiff}{2.9}
\newcommand{\HIIlocalAvN}{98}
\newcommand{\HIIlocalAvtopthirdf}{1.1}

\newcommand{\KDohtwoVSZninefourstd}{0.07}

\newcommand{\NohsixNtwoVSZninefourmed}{-0.10}
\newcommand{\NohsixNtwoVSZninefourstd}{0.21}

\newcommand{\NohsixOthreeNtwoVSNohsixNtwostd}{0.12}

\newcommand{\PTohfiveVSZninefourmed}{-0.50}

\newcommand{\PTohfiveVSNohsixNtwostd}{0.30}

\newcommand{\NHrms}{0.11}
\newcommand{\IDnewpneightfournine}{HII153}

\newcommand{\IDnewpneighteightnine}{HII166}
\newcommand{\IDpnntwozerofiveonezeroonenine}{PN130}

\newcommand{\newpneighteightnineNohsixNtwo}{8.94\pm0.02}

\newcommand{\newpneightfournineNohsixNtwo}{8.36\pm0.04}

\newcommand{\favsepMIN}{1.93}
\newcommand{\favsepKPC}{0.4}
\newcommand{\favsepR}{0.84}
\newcommand{\favsepVEL}{24}
\newcommand{\ZKHgradpKPC}{11-26}
\newcommand{\ZKHgradpRHO}{16-32}

\def\arcmin{\char'023 }
\def\arcsec{\char'175 }
\def\asec{\char'175 }

\begin{document}

\title{The metallicity profile of M31 from spectroscopy of hundreds of HII regions and PNe}
\author{Nathan E. Sanders, Nelson Caldwell, Jonathan McDowell}
\affil{Harvard-Smithsonian Center for Astrophysics \\ 
       60 Garden Street, Cambridge, MA 02138, USA}
\email{nsanders@cfa.harvard.edu} 
\author{Paul Harding}
\affil{Department of Astronomy, Case Western Reserve University \\
 Cleveland, OH 44106-7215, USA}

\begin{abstract}

The oxygen abundance gradients among nebular emission line regions in spiral galaxies have been used as important constraints for models of chemical evolution.  We present the largest-ever full-wavelength optical spectroscopic sample of emission line nebulae in a spiral galaxy (M31).  We have collected spectra of \CountHiiTotal{} HII regions and \CountPNTotal{} planetary nebulae with the Hectospec multi-fiber spectrograph of the MMT.  We measure the line-of-sight extinction for \HIINAv\ HII regions and \PNNAv\ PNe; we derive oxygen abundance directly, based on the electron temperature, for \PNNdirect\ PNe; and we use strong line methods to estimate oxygen abundance for \HIINNohsixNtwo\ HII regions and nitrogen abundance for \HIINPVTNHONS\ HII regions.  The relatively shallow oxygen abundance gradient of the more extended HII regions in our sample is generally in agreement with the result of \cite{zaritsky94}, based on only 19 M31 HII regions, but varies with the strong-line diagnostic employed.  Our large sample size demonstrates that there is significant intrinsic scatter around this abundance gradient, as much as $\sim3$~times the systematic uncertainty in the strong line diagnostics.  The intrinsic scatter is similar in the nitrogen abundances, although the gradient is significantly steeper.  On small scales (deprojected distance $<0.5$ kpc), HII regions exhibit local variations in oxygen abundance that are larger than \HIIlocalOHtopthirdf{}~dex in $33\%$ of neighboring pairs.  We do not identify a significant oxygen abundance gradient among PNe, but we do find a significant gradient in the [\ion{N}{2}] ratio that varies systematically with surface brightness.  Our results underscore the complex and inhomogeneous nature of the ISM of M31, and our dataset illustrates systematic effects relevant to future studies of the metallicity gradients in nearby spiral galaxies.

\end{abstract}

\keywords{Catalogs --- Galaxies: individual (M~31) --- Galaxies: evolution ---  Galaxies: abundances --- HII regions: general --- Planetary nebulae: general}

\section{INTRODUCTION}
\label{sec:intro}

The galactocentric radial gradient of chemical abundance within spiral galaxies has become an important parameter in modeling the chemical evolution of galaxies \citep{henry99}.  The gradient is the manifestation of a variety of physical processes acting from galaxy formation to the present, including gas infall, star formation history, stellar initial mass function, and radial migration.   Historically, the observation of abundance gradients has motivated the development of analytic models of chemical evolution \citep[][and references therein]{lb75} and served as a valuable constraint for detailed modeling \citep{maciel04,molla05,Scannapieco08,mag-stan09,Carigi10}.  Moreover, the characterization of the abundance profiles of spiral galaxies has significant applications to other problems, including the interpretation of stellar populations \citep{Massey98} and the study of supernova explosion sites \citep{Levesque10bh}.

The oxygen abundance of gas throughout the disk of star-forming galaxies can be measured via optical spectroscopy.   O is the most easily accessible metallicity proxy due to its high relative abundance and its strong optical emission lines from both its major ionization states (O$^+$ and O$^{++}$).  A variety of diagnostics have been developed to estimate O abundance from the flux ratios of prominent optical emission lines \citep[see e.g.][]{stas02,KE08,Lopez2010}.  These include "direct" methods, whereby the electron temperature of the nebula is derived from measurement of the weak auroral line [\ion{O}{3}]~$\lambda4363$ \citep{oandf,Garnett92}, and  "statistical" or "strong line" measures, where the abundance is inferred from ratios of only the brightest elemental lines.  Different abundance diagnostics carry systematic discrepancies as large as $0.7$~dex (a factor of 5), which must be carefully considered when interpreting results \citep{KE08}.

Strong line methods have enabled extensive observational studies of the abundance distribution of distant spiral galaxies.  While high quality spectra can be obtained for the nearby HII regions of the Milky Way, galactocentric distance can be determined more easily for external galaxies \citep{stanghellini08,Henry10}.  \cite{shields78} derived a negative (outward-decreasing) abundance gradient from just three HII regions in M101, and by this time it was already expected from theory that negative radial abundance gradients would be characteristic of all spiral galaxies.  \cite{zaritsky94} compiled radial O abundance profiles for 39 spiral galaxies, finding gradients ranging from $0$ to $-0.23$~dex~kpc$^{-1}$---only the abundance profile of NGC 2541 was inconsistent with a negative or flat gradient.  

Because PNe abundances should reflect the older ISM of their progenitors, differences between the HII region and PNe abundance gradients can be used to infer time-variation in the radial abundance trend of a galaxy \citep[e.g.][for M33]{mag-stan09}.  A sample of high-luminosity PNe should reflect current oxygen abundances in a galaxy, by selecting massive progenitors which would have formed recently.  The oxygen abundances in a sample containing only the brightest PNe in a galaxy would represent populations with ages from $\sim 3\times10^7-10^{10}$ years earlier than the HII regions in the same galaxy \citep{mag-stan09}.  For surveys penetrating more deeply, the proportion of PNe from less massive progenitors will grow quickly both because less massive stars are more common and because their resulting PNe are longer lasting.  However, stars of mass $\lesssim1.5~\mbox{M}_{\sun}$ may not form PNe because their envelope is ejected so slowly as to disperse before being ionized \citep{stas02}.

While $\sim10^3$ bright HII regions ($L_{\rm H\alpha}\gtrsim5\times10^{34}$~ergs~s$^{-1}$) are known in M31 \citep{baade64,pellet78,WB92,Azimlu11}, previous spectroscopic surveys to determine abundance have provided abundance estimates for $\lesssim 7\%$ of them.  \cite{blair82} found a significant radial abundance gradient in a sample of 11 HII regions: O/H decreases by a factor of 4 from about $4 - 23$~kpc in galactocentric distance.  This gradient is in agreement with an earlier study of 8 HII regions in M31 by \cite{dennefeld81}.  In an analysis of the data from both \cite{blair82} and \cite{dennefeld81}, \cite{zaritsky94} quantified the oxygen abundance gradient as relatively shallow among their sample of galaxies: $-0.020 \pm 0.007$~dex/kpc\footnote{Hereafter, we quote physical values for M31 based on a distance of 770~kpc \citep{freedman90}.  \cite{zaritsky94} used an earlier distance measurement of 700~kpc, so we apply a $10\%$ correction to their measured abundance gradient.}. In comparison, the largest gradient in the \cite{zaritsky94} sample was $-0.231 \pm 0.022$~dex/kpc (NGC 3344). Most recently, \cite{Galarza99} surveyed 46 HII regions out to about 20~kpc in the Northeastern portion of M31, finding a radial gradient in $R_{23}$ that is consistent with \cite{blair82}, but only among objects whose morphology was classified as "center-brightened" by \cite{WB92}.  Among HII regions of other morphologies, they report a flat abundance profile.   Each of these previous surveys has relied on various strong line abundance diagnostics.  Additionally, abundance gradients have been derived from surveys of M31's stellar population.  \cite{Trundle02} found a flat radial metallicity profile among seven young B stars.  \cite{worthey05} estimated abundances for an older population, red giant stars, from color-magnitude diagrams of 11 fields in M31.  For radii $<25$~kpc, they find a negative gradient similar to the nebular result of \cite{dennefeld81} and \cite{blair82}, but they report a flattening at larger radii.

M31 hosts as many as $10^4$~PNe, about twice as many as the Milky Way \citep{nolthenius87}.  In a kinematic survey, \cite{merrett06} cataloged 2615 of M31's PNe.  Abundances have been previously derived for less than 1\% of these.  \cite{jacoby86} determined abundances for three PNe: two in the halo and one in the outer disk.  The results of the largest previous spectroscopic survey of M31 PNe, including 70 objects, have not yet been published \citep{Kniazev05}.  A survey of 30 nebulae in the bulge of the galaxy was performed by \cite{richer99}, but only a lower limit of abundance could be derived for 14 of the nebulae.  Because the survey only extended out to $\lesssim4$~kpc and because only the bulge population is sampled, \cite{richer99} did not investigate the radial abundance gradient.  Recently, \cite{Kwitter12} have derived abundances for 16 PNe in the outer disk of M31.

By providing a more thorough characterization of the abundance profile of in M31, this paper seeks to enable an improved understanding of the chemical evolution of M31 and similar spiral galaxies.  In Tables \ref{tab:obj}-\ref{tab:derPN} we present the largest available spectroscopic catalogs of HII regions and PNe in M31. Our observations come from the Hectospec multifiber spectrograph on the MMT, whose multiplexing ability provides a large advantage over previous surveys of these objects.  In Section~\ref{sec:observations} we describe the observational parameters and analytical techniques used to produce the catalog.  In Section~\ref{sec:disc} we compare our results to previous publications and discuss trends and implications identified in the catalog.  Our principal findings are that a significant negative abundance gradient is only demonstrated among the brightest and most diffuse HII regions.  The radial profile is more flat among dimmer or more compact HII regions, and in general there is a large amount of scatter in the physical properties of the ISM of M31.  We characterize this scatter in terms of the radial distribution of extinction and abundance in HII regions and PNe, and also in terms of the discrepancies among neighboring objects.  We provide a summary of major results in Section~\ref{sec:conc}.\section{OBSERVATIONS}
\label{sec:observations}

\subsection{Data collection}
\label{sec:obs-dcol}

Small, resolved objects and unresolved H$\alpha$ features  were selected  as HII region candidates by inspecting the images of the Local Group Galaxies Survey \citep[LGGS,][]{LGGSHa}. Additionally, some objects observed as part of the M31 cluster survey of \cite{caldwell09} had strong emission, and are included in the present study. Some objects from the planetary nebula catalog of \cite{merrett06} as well as strong and unresolved [\ion{O}{3}] features from the LGGS images were observed as PN candidates.  We have excluded from our sample any objects identified as emission line stars by \cite{LGGSHa}, one object which was found to have broad emission lines characteristic of supernova remnants, and $\sim20$ objects which showed broad emission features characteristic of WR stars.  

We note that many M31 HII regions have diameters greater than 50\arcsec\ \citep{arp73}, far too large to be encompassed by a single Hectospec fiber, which subtends 1.5\arcsec on the sky ($\sim5$ pc at the distance of M31).  Moreover, to facilitate a future study of the kinematic properties of nebulae in M31, we attempted to avoid HII regions which would have a large internal velocity dispersion.  For both these reasons, the largest HII regions were therefore intentionally omitted from the sample, although some objects remain (mostly the star clusters) that were estimated by \cite{blair82} as at least 48~pc in diameter.  We assume that any inhomogeneities in spectral properties within each HII region are small and random and therefore that the region sampled by the Hectospec fiber is representative of the whole \citep[see e.g.][]{mccall85,Pellegrini10}.

Optical spectra were obtained with the Hectospec multi-fiber positioner and spectrograph on the 6.5m MMT telescope \citep{fabricant05}.  The Hectospec 270 gpm grating was used and provided spectral coverage from $3650-9200${\AA} at a resolution of $\sim5${\AA}.  Some spectra did not cover [\ion{O}{2}]$\lambda 3727$, because of the design of the spectrograph (alternate fibers are shifted by 30{\AA}), and the small blueshift of M31.  The observations were made in the period from $2004-2011$ as a component of a survey previously described in \cite{caldwell09} and were reduced in the uniform manner outlined there.  The frames were first debiased and flat fielded.  Individual spectra were then  extracted and wavelength calibrated.  Standard star spectra obtained intermittently were used for flux calibration and instrumental response.  Sky subtraction is achieved with Hectospec by averaging spectra from ``blank sky'' fibers from the same exposures or by offsetting the telescope by a few arcseconds (see \citealt{caldwell09}).  Because local background subtraction could potentially subtract object flux from extended HII regions, we compare the spectra reduced from repeated observations of the same objects using different sky spectra, local and distant.   For both stellar and diffuse objects, we find only  small differences in the ratios of H$\alpha$/H$\beta$ and also [\ion{N}{2}]/H$\alpha$.  The rms of the discrepancy in the log of the flux ratio is $\sim0.05$~dex and the mean difference is $\sim0.01$~dex.  We therefore conclude that the sky subtraction is adequate.

\begin{figure}
\centering
 	\plotone{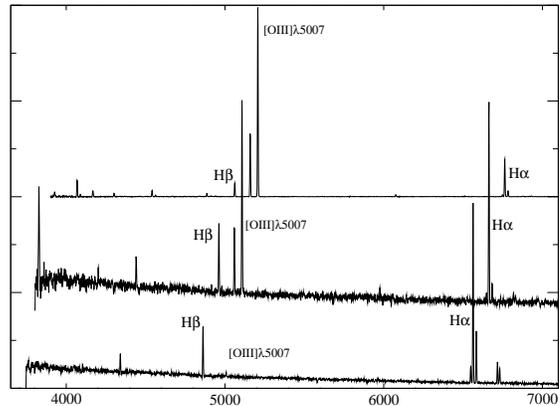}
\caption{Three spectra from Hectospec. From the bottom, a high-metallicity HII region (object \IDnewpneighteightnine{}), a low-metallicity HII region (object \IDnewpneightfournine{}), and a PNe (object \IDpnntwozerofiveonezeroonenine{}).  The two HII regions are separated by only \favsepMIN{}\arcmin\ on the sky, corresponding to a separation of $\sim\favsepKPC{}$ kpc at the distance of M31.  For clarity, the relative flux scales were set such that the strongest emission line in each spectrum (H$\alpha$ for HII regions and [\ion{O}{3}]$\lambda 5007$ for the PN) are the same height and the wavelength zero point was offset by 100 and 200\AA \ for the two upper spectra. [\ion{O}{2}]$\lambda 3727$ falls off of the original spectrum for the upper and lower spectra. }\label{fig:spec}
\end{figure}

\begin{figure*}
\centering
        \plotone{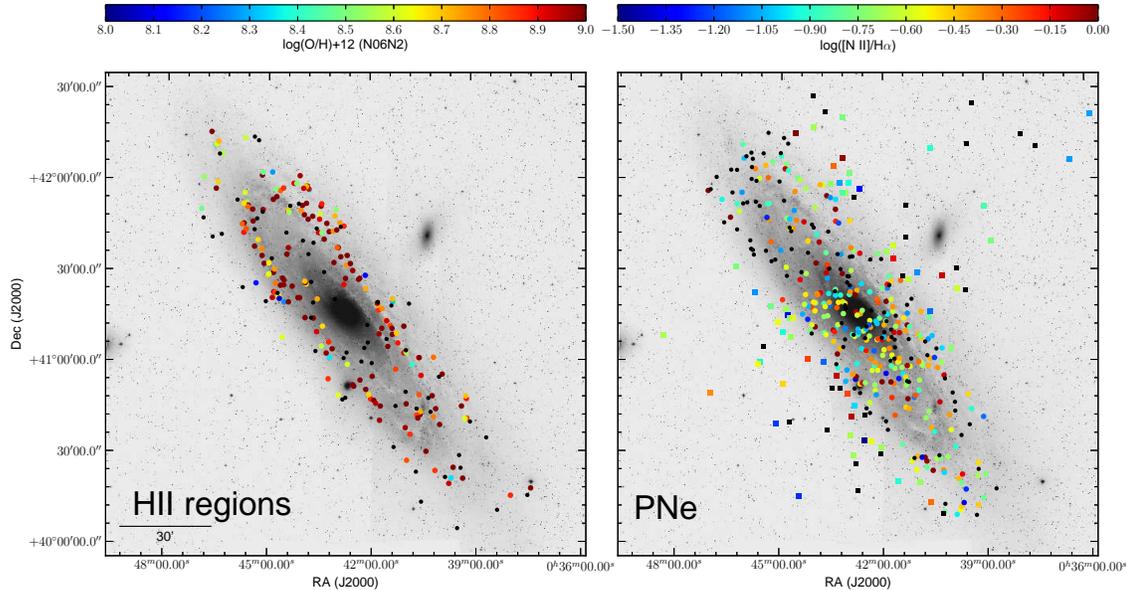}
\caption{Locations of M31 HII regions and PNe observed with Hectospec, overlaid on the M31 mosaic from the Digitized Sky Survey (DSS).  The HII region symbols are colored to oxygen abundance. using the N06 N2 diagnostic, and the PNe are colored according to their N2 flux ratio (see Section~\ref{sec:obs-abundance}).  Halo PNe are denoted with squares and objects whose abundance/flux ratio could not be measured are marked in black.}\label{fig:bothfinder}
\end{figure*}

Each of 25 1~degree fields in M31 was exposed for between 1800 and 4800~s.  The spectra of objects that were observed multiple times (in overlapping fields) have been combined, effectively summing those integration times. Sample spectra are shown in Figure \ref{fig:spec}.
The locations of emission nebulae included in this study are shown in Figure \ref{fig:bothfinder}.     Many of our HII regions fall in the ``Ring of Fire,'' a circular feature visible in HI density maps extending from about 8 to 15 kpc in the disk of M31 \citep{Sofue81}. 

Velocities were measured using the SAO xcsao software and emission line templates (one typical of HII regions and another of PNe).  Repeat measurements of 114 objects gave an rms of a single measurement of 2.1 km~s$^{-1}$.  We also compared our velocities with the work of \cite{Halliday06}, who also used a multifiber spectrograph, and \cite{merrett06} who used the Planetary Nebula Spectrograph (which imaged in [\ion{O}{3}]).  The \cite{Halliday06} comparison revealed a mean offset of +7.1 km~s$^{-1}$ with an rms of 5.4 km~s$^{-1}$, while the \cite{merrett06} comparison resulted in a mean offset of +2.9 km~s$^{-1}$ with an rms of 15.6 km~s$^{-1}$. The \cite{merrett06} rms was expected to be larger because of their use of the single emission line. These comparisons led us to estimate our mean error to be 3  km~s$^{-1}$. The final step in the reduction was the transformation of the spectra to zero velocity using the observed velocities.

\subsection{Classification}
\label{sec:obs-class}

\begin{figure}
\centering
 	\plotone{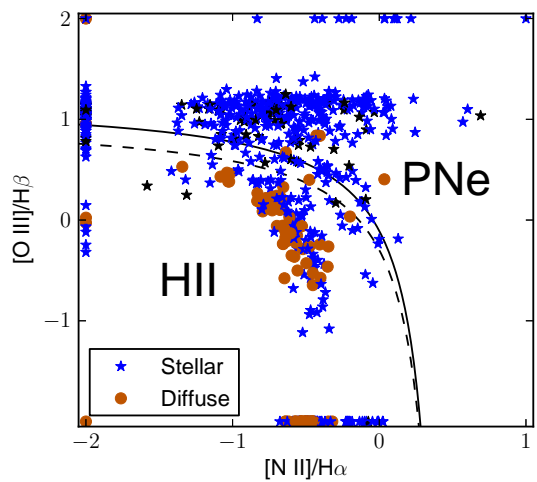}
\caption{Excitation mechanism diagnostic diagram \citep{Baldwin81} distinguishing PNe from HII regions for M31 emission line nebulae observed in this study.  Emission line flux ratio error bars are typically smaller than the points.  The solid black classification divider is from \cite{Kniavez08}; the dashed line is our amended divider, on which our spectroscopic classification is based.  The data points are color coded according to H$\alpha$ morphology (stellar or diffuse); objects identified as being in the halo based on their position along the minor axis ($|Y| > 4$ kpc) are colored black.  Objects shown on the edge of the figure denote flux ratio limits for lines which were not detected.}\label{fig:bpt}
\end{figure}

These candidates were formally classified according to two-dimensional emission line ratio tests of excitation mechanism \citep[BPT diagrams,][]{Baldwin81}.  Specifically, we apply equation 1 of \citep{Kniavez08}, which distinguishes HII regions from PNe based on their locations in a diagram of [\ion{O}{3}]/H$\beta$ (O3) versus [\ion{N}{2}]/H$\alpha$ (N2) (Figure \ref{fig:bpt}).  These tests have the advantage of relying on ratios of strong emission lines that are near to each other in wavelength and therefore not sensitive to reddening corrections or instrumental response.  The measurement of line fluxes is described in Section~\ref{sec:obs-lineflux}.  We adopt a slight amendment to the dividing line proposed by \citep{Kniavez08}, illustrated by the dashed line in Figure \ref{fig:bpt}.  This amendment is favored because it classifies as PNe a number of objects that we have reason to believe are not HII regions.  These objects have stellar morphologies (see below) and many of them are in the halo, based on their position along the minor axis ($|Y| > 4$ kpc), as is illustrated in Figure \ref{fig:bpt}.  It is not surprising that there may be small offsets in the appropriate line ratio diagnostics for different samples given slight differences in flux-measurement methodology.  The amended classification we adopt to distinguish HII regions from PNe is: 

\begin{equation}
{\rm O3}>(0.61/({\rm N2}-0.47))+1.0
\end{equation}

In total, we identify \CountPNTotal{} PNe and \CountHiiTotal{} HII regions among our spectroscopic sample.  Of these objects, \CountPNADC{} PNe and \CountHiiADC{} HII region spectra were observed with the atmospheric dispersion compensator \citep[ADC,][]{Fabricant08}.  For some spectra, the ADC malfunctioned, and for those we do not present quantities derived from emission line flux ratios with large wavelength separations.  \CountPNhalo{} of the PNe do not appear to be projected onto the disk of M31 (according to the distance along the minor axis, $|Y| > 4$ kpc, see Section~\ref{sec:obs-galdist}), and we therefore assume they are associated with the halo population.  \CountPNNGC{} of the PNe are hosted by the dwarf satellite galaxy \object{NGC 205} rather than in M31, according to their position and velocity in the kinematic survey of \cite{merrett06}.  Because the S/N of these spectra are such that it is not possible to derive their abundances, we remove them from the sample and do not discuss them further.  An additional \CountNoClass{} objects could not be classified because the S/N in their [\ion{O}{3}] and H$\beta$ emission lines were too weak.  We do not discuss these unclassified objects further.

In addition to this spectroscopic classification, we have made a morphological classification for each object based on its appearance in the LGGS H$\alpha$ images.  Objects whose H$\alpha$ emission appears unresolved at the resolution of these images ($\sim1\asec$) are classified as ``stellar'' and those that appear extended and nebulous are classified as ``diffuse.''  We note that this is similar to the classification scheme of \cite{WB92} referred to by \cite{Galarza99}, where their type ``C'' or ``center-brightened'' corresponds to our ``stellar'' and their type ``D'' or ``diffuse'' corresponds to the same.  We find that \diffusepne{} spectroscopically-classified PNe are diffuse.  Because most of these objects are near the dividing line in the BPT diagram, it is difficult to determine whether these are PNe embedded in HII regions or simply poor spectroscopic classifications.  However, these only amount to $\sim\diffusepneper{}$\% of our full sample of PNe.

About half of our spectroscopically-classified HII regions are stellar.  From Figure \ref{fig:bpt}, it is clear that the stellar HII regions are more likely to have high O3 or N2 ratios relative to diffuse HII regions and are therefore nearer to the dividing line in the BPT diagram.  It is possible that some of these objects are PNe with an unusually low value of [\ion{O}{3}]/H$\beta$.  Among the sample of Galactic PNe in \cite{Henry10}, there are some objects (M1-11 and M1-12) that fall on a similar region of the BPT diagram.  According to the Catalogue of Galactic Planetary Nebulae \citep{PNcat}, these objects were originally classified as PNe by \cite{Minkowski46}, whose classifications were essentially morphological.  It is also possible that these are extended HII regions whose H$\alpha$ emission is simply dominated by the emission in the immediate vicinity of one bright star, and therefore appear compact on the LGSS images.  Therefore the diffuse subset of HII regions in our sample may represent a cleaner sample, less likely to have PN contamination.

\subsection{Galactocentric distance}
\label{sec:obs-galdist}

The deprojected galactocentric distance was calculated following \cite{haud81}.  We assume a distance to M31 of 770~kpc \citep{freedman90} and an inclination of $i=12.5^\circ$ \citep{simien78}.  The adopted coordinates of the galactic center and position angle of the major axis, precessed from \cite{haud81}, are:

\begin{equation*}
\alpha_0 =  00^\circ42'44''.52 \qquad \mbox{(J2000)} \\
\end{equation*}
\begin{equation*}
\delta_0 = +41^\circ16'08''.69 \qquad \mbox{(J2000)} \\
\end{equation*}
\begin{equation*}
\phi_0 = 37^\circ 42' 54''
\end{equation*}

Hereafter, we use ``galactocentric distance'' to refer to this deprojected distance.  Any PNe with a distance projected along the semiminor axis ($Y$) greater than 4 kpc, we associate with the halo population and do not consider in our sample when fitting for radial trends in the disk.  Additionally, there are 3 objects that we consider halo PNe given their projected distance $|Y|>5$~kpc, despite the fact that they fall in the HII region regime of the BPT diagram.

\subsection{Emission line fluxes}
\label{sec:obs-lineflux}

The line fluxes were measured using the line profile-fitting capabilities provided by the IRAF package \texttt{fitprofs}.  We fit Gaussian profiles to a wavelength range $20$ \AA{} in width centered on the rest frame wavelength of each line.  We fit a linear continuum to $20$ \AA{} regions of the spectra off the wings of each line.  For groups of nearby lines likely to be blended, we fit simultaneous Gaussians.  The line flux is estimated as the integral of the fitted profile with the continuum subtracted.  We estimate the uncertainty in the line fluxes using the Monte Carlo methodology implemented by \texttt{fitprofs}.  In these Monte Carlo simulations, random noise is repeatedly added to the spectrum according to a simple linear noise model dependent on two parameters: the gain of the CCD and a noise floor.  We estimate the gain and noise floor for each individual line profile by comparing the spectrum of the background regions to the corresponding variance spectrum.  We have compared this estimate of the line flux uncertainty to the discrepancy between line flux ratios measured in repeated observations of the same objects and find them to be comparable; for the H$\alpha$/H$\beta$ ratio, the Monte Carlo uncertainty estimate and the discrepancy among the repeated observations are each $\sim3\%$ in the mean.  Based on the Monte Carlo estimate of the uncertainty in the line flux, we require that S/N$>3$ and otherwise report a non-detection.  Additionally, we record the equivalent width as measured by \texttt{fitprofs}, if the sky-subtracted continuum level is more than $2\sigma$ greater than zero (where the continuum level and its standard deviation, $\sigma$, are measured from the background regions of each line).

The [\ion{O}{2}] $\lambda\lambda3726-3729$ doublet is not resolved in these spectra, and the sum effectively measured will henceforth be referred to as [\ion{O}{2}] $\lambda3727$.  Moreover, because $\lambda3727$ is on the very edge of our spectral range, it has a large flux-calibration uncertainty associated with it, which is propagated through to our estimate of the flux uncertainty.  As mentioned above, for some spectra, $\lambda3727$ is outside of the observed spectral range.

We note that we have made no correction for underlying stellar absorption, but we estimate that this will not significantly affect the measurement of the Balmer line fluxes. Using models from Starburst99 \citep{Starburst99}, we subtracted off underlying continua for populations with ages ranging from 4 to 20 Myr for a representative low- and high-metallicity HII region (objects \IDnewpneighteightnine{} and \IDnewpneightfournine{}), see Figure \ref{fig:spec}). We then measured H$\gamma$ and H$\beta$.  The ratios of those two lines changes at worst by $1\%$ from the uncorrected values (the worst case results from assuming the youngest age for the underlying population).  As a further test of the small effect that the underlying continuum has on our measurements, we also measured H$\alpha$ equivalent widths and compared those with $A_v$ (derived from the H$\alpha$/H$\beta$ emission line ratio), and found no significant correlation. This indicates in general that Balmer absorption doesn't significantly affect the measured Balmer ratios for the objects in our sample.

We present the measured line fluxes and their uncertainties for each object in Table \ref{tab:mea}.  For all parameters derived from these line fluxes, we propagate the uncertainty in the line flux via Monte Carlo simulations.  In these simulations, we sample from a Gaussian line flux probability distribution with a mean and standard deviation as reported in Table \ref{tab:mea}.  We then report the median and standard deviation of the resulting distribution as our best estimate of the derived parameters and its uncertainty.

We characterize the surface brightness of each emission line region based on its observed H$\alpha$ emission line flux, even if the objects are unresolved.  H$\alpha$ is used for this purpose because it is strong and easily detected in nearly every spectrum.  
However, the 1.5\arcsec\ Hectospec fibers generally do not cover the entire area of the diffuse emission nebulae.  Moreover, as the observation nights were not all photometric, there could be photometric uncertainties of a factor of a few.  We attempted to place all the spectra on the same photometric scale by using the multiple observations, and find that this was successful to within a factor of 1.3.  

\begin{figure}
\centering
 	\plotone{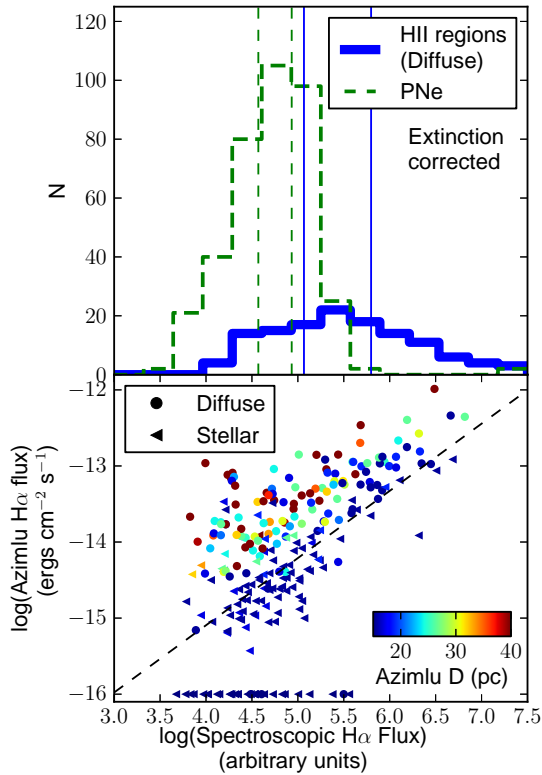}
\caption{Top: The surface brightness distribution of diffuse HII regions (blue) and disk PNe (green), as measured spectroscopically from the extinction-corrected H$\alpha$ emission line flux (arbitrary units).  The solid and dashed vertical lines denote the surface brightness limits which split the sample into three equally sized bins of HII regions and PNe, respectively.  Bottom: Comparison of spectroscopically-measured H$\alpha$ fluxes (not extinction corrected) and the photometric catalog of \cite{Azimlu11} for HII regions only, with color coding by the catalog diameter.  Symbol shape denotes our morphological types.  Points on the edge of the figure do not have matches within 5\asec.  The dashed line is the best fit among small ($D<20$~pc) objects.}\label{fig:fluxdens}
\end{figure}

We therefore divide our sample into three surface brightness classes (1: ``dim,'' 2: ``normal'', 3: ``bright''; see Figure~\ref{fig:fluxdens}) based on the extinction-corrected H$\alpha$ line flux, rather than assert a precise flux measurement.  We establish separate brightness classes for the HII regions (based on the diffuse subset only) and PNe.  For spectra where the H$\beta$ line is not detected, we assume the median $A_V$ for that class (HII region or PNe) for the purpose of calculating the extinction corrected H$\alpha$ line flux.  The diffuse HII regions in our sample are at $\sim\SBmeanRatio$ times higher surface brightness than the PNe, in the median (or $\sim\SBmeanRatioC$ times brighter after extinction correction).  Moreover, the HII regions encompass a larger range in surface brightness, extending to nearly two orders of magnitude brighter in H$\alpha$.

We have estimated the apparent brightness limits corresponding to these surface brightness classes by matching the HII regions in our sample to the catalog of \cite{Azimlu11}, using the nearest objects within 5\asec.   They use an automated code to segment diffuse emission in continuum-subtracted H$\alpha$images of M31 for the identification of HII regions and PNe and to measure the H$\alpha$ flux and diameter.  Figure~\ref{fig:fluxdens} shows the relation between our spectroscopic H$\alpha$ flux measurements and the photometry from \cite{Azimlu11}.  The scatter in the relation is correlated with the HII region diameter, because HII regions which \cite{Azimlu11} segment as larger objects are have a correspondingly smaller fraction of their flux fall on our Hectospec fibers.  Among small objects ($D<16$~pc), the scatter is $\sim\AzimluCompScat$~dex.  This relation implies that our HII~region flux bins have edges at $\sim[\AzimluFHIIa,\AzimluFHIIb]\times10^{-15}~{\rm ergs}~{\rm cm}^{-2}~{\rm s}^{-1}$ and, for PNe, $\sim[\AzimluFPNa,\AzimluFPNb]\times10^{-15}~{\rm ergs}~{\rm cm}^{-2}~{\rm s}^{-1}$.

\subsection{Extinction}
\label{sec:obs-extinction}

A reddening correction was applied to restore the Balmer recombination decrement of each spectrum to its theoretical value. We assume $\mbox{H}\alpha/\mbox{H}\beta$=2.85, which corresponds to T=10,000K and $n_e=10^4\mbox{cm}^{-3}$ for Case B  recombination.  However, the Balmer ratios are not very sensitive to any of these parameters: the assumption of Case A  only alters the ratio by $\lesssim0.05$ and extreme temperatures and densities also have little effect: $\mbox{H}\alpha/\mbox{H}\beta=3.04$ for $T=5,000$~K and $n_e=10^2~\mbox{cm}^{-3}$, $\mbox{H}\alpha/\mbox{H}\beta=2.73$ for $T=20,000$~K and $n_e=10^6~\mbox{cm}^{-3}$ \citep{oandf}.  The extinction curve given in \cite{cardelli89} was applied, with a value of 3.1 adopted for $R_V$.  In cases where we derive a negative value of the extinction in the visual band ($A_V<0$), we instead assume $A_V=0$.

The value of the extinction derived for each object is presented in Tables~\ref{tab:derHII} and \ref{tab:derPN}.

\subsection{Direct abundance estimation}
\label{sec:obs-abundanceD}

We derive ``direct'' oxygen abundance estimates for PNe by estimating the electron temperature of the gas' dominant excitation zone, which can be done only if a temperature-sensitive line is detected.  We use [\ion{O}{3}] $\lambda4363$ exclusively for this purpose; although auroral lines from other ions  (e.g., [\ion{N}{2}] $\lambda5755$) are detected in a few spectra, their numbers are not sufficient for a statistical sample and their S/N is typically much lower than [\ion{O}{3}] $\lambda4363$.  This prescription is also applied to our HII regions, however direct abundances can be derived for only $\HIINdirect$ HII regions due to the weakness of the [\ion{O}{3}] $\lambda4363$ line.

The electron temperatures are estimated using IRAF's five-level nebular modeling package \texttt{nebular} \citep{shaw94}.  The \texttt{nebular} task \texttt{temden} is first applied to iteratively estimate the O$^{++}$ temperature ($T_e(\rm O^{++})$) and density ($n_e$) of the nebula from the [\ion{O}{3}] and [\ion{S}{2}] line ratios, respectively.  When the [\ion{S}{2}] lines are note available, we assume a reasonable range of densities, $n_e=15\pm2\times10^3~{\rm cm}^{-3}$ \citep[see e.g.][]{Kwitter12}.  If the measured line ratios correspond to unphysical conditions (outside the range for which \texttt{temden} is calibrated, $500<T_e(\rm O^{++})<10^5$~K and $1<n_e<10^8$~cm$^{-3}$), we do not calculate the direct abundance.  The O$^+$ temperature is then estimated using the linear empirical relation of \cite{Garnett92}.  The O$^{+}$ and O$^{++}$ abundances are then estimated using the density, ionic temperatures, and [\ion{O}{2}] and [\ion{O}{3}] line ratios following the ionization correction factor (ICF) prescription of \cite{Shi06}.  The total oxygen abundance is taken to be the sum of these two ionic abundances.  This methodology is similar to that applied in, for example, \cite{Bresolin10}.  

A variety of studies have shown that stellar evolution of PNe progenitors should not modify oxygen abundance at a level significant for the identification of abundance gradients from the nebulae \citep[][and references therein]{richer07}.  Furthermore, it has been shown that bright PNe (within $\sim2$ mag of the brightest in the galaxy), presumably from more massive stars with shorter lifetimes, have approximately the same oxygen abundances as the surrounding ISM and HII regions, to within the observational uncertainty \citep{richer07}.  It has been demonstrated that O, as well as Ar and Ne, abundance gradients in the Milky Way as measured with HII regions are reflected in observations of PNe \citep{pottasch06}.  However, oxygen may be dredged up in some low-metallicity cases (log(O/H)$+12 \lesssim 8$) and thereby enrich the PNe relative to its progenitor by $\triangle $log(O/H)$ \lesssim 0.3$ dex \citep{richer07,hernandez09,mag-gon09}.  In high-mass PNe, oxygen may be depleted by $\triangle $log(O/H)$ \lesssim 0.1$ dex during the lifetime of the progenitor via the ON-cycle \citep{hernandez09}.  Moreover, it has been observed that the N abundance of PNe will often exceed that of the local ISM \citep{richer07,hernandez09}---justified theoretically by nitrogen production via the CN or ON cycle during the first and second dredge ups.  \cite{Henry10} have established empirically that the oxygen abundance gradient of Milky Way PNe does not depend on the Peimbert type of the PNe or the vertical distance from the plane of the Galaxy; we therefore do not consider these parameters here.

\subsection{Strong line diagnostics}
\label{sec:obs-abundance}

HII region abundances were calculated from the extinction-corrected line flux ratios according to a number of independently calibrated abundance diagnostics from the literature.  These diagnostics each depend on a different combination of emission line ratios.  Each calibration has its own characteristic scatter and systematic offset as compared to the other methods \citep{KE08}.  It is unclear that any particular method is superior, and some methods may be more applicable than others for certain comparisons.  In particular, diagnostics tied to the direct abundance scale may agree better with stellar oxygen abundances within the same galaxy \citep[see e.g.][]{Bresolin09}.  We therefore employ multiple methods and keep their differences in mind when discussing results.  We report the abundance derived from each method in Tables~\ref{tab:derHII}  (HII regions) and \ref{tab:derPN} (PNe).

First, we apply the $R_{23}$ oxygen abundance calibration of \cite{zaritsky94}, an average of three earlier methods, hereafter referred to as ``Z94.''  Z94 is only calibrated for the higher-metallicity upper branch of the well-known $R_{23}$-abundance degeneracy.  The majority of M31 HII regions may be expected to fall on this upper branch, given that all the M31 HII regions in the compilation of \cite{zaritsky94} did.  If the measured line ratios correspond to an abundance outside of the range for which Z94 is calibrated (generously, $8.4<$log(O/H)$+12<9.6$), we do not record the measurement.

Second, we apply the [\ion{N}{2}]/[\ion{O}{2}] oxygen abundance calibration of \cite{kewley02}, hereafter referred to as ``KD02.''  \cite{kewley02} synthesize a variety of modern photoionization models and observational calibrations to produce recommendations for producing an abundance estimate given different permutations of available emission lines.  We implement the prescription outlined in the appendix of \cite{KE08}, as follows.  We use the [\ion{N}{2}]/[\ion{O}{2}] ratio to break the degeneracy between the upper and lower branches of $R_{23}$.  For the upper branch, we employ the [\ion{N}{2}]/[\ion{O}{2}] calibration of \cite{kewley02}.  For the rare lower branch cases, we average the $R_{23}$ diagnostics of \cite{mcgaugh91} and \cite{KK04}.  If the measured line ratios correspond to an abundance outside of the range for which KD02 is calibrated ($8.2<$log(O/H)$+12<9.6$), we do not record the measurement.

Third, we apply the empirical [\ion{N}{2}]/H$\alpha$ (``N2'') and [\ion{O}{3}]/[\ion{N}{2}] (``O3N2'')  oxygen abundance calibrations of \cite{Nagao06}, hereafter referred to as ``N06.''  We prefer the N06 diagnostic to the similar ``PP04'' N2 and O3N2 diagnostics of \cite{PP04} because N06 is well-calibrated in the high-metallicity regime of M31 using data from the SDSS galaxies \citep{T04}.  If the measured line ratios correspond to an abundance outside of the range for which N06 is calibrated ($7.0<$log(O/H)$+12<9.5$), we do not record the measurement.  Because this diagnostic relies on the \ion{N}{2} lines to measure the O abundance, scatter is introduced by variations in the N/O ratio \citep{PMC09}.  We note also that the O3N2 diagnostic is not reliable when ${\rm O3N2}\gtrsim100$, due to line saturation, but this only occurs in a metallicity regime lower than that sampled here (log(O/H)$+12\lesssim7.5$; \citealt{PP04,Nagao06}).

Fourth, we apply the excitation parameter (``$P$ method'') oxygen abundance calibration of \cite{PT05}, hereafter referred to as ``PT05.''  This is an updated version of the calibration first defined in \cite{P01} (P01).  $P$ is calculated from the ratio of [\ion{O}{3}] to ([\ion{O}{2}]+[\ion{O}{3}]).  PT05 additionally relies on the $R_{23}$ line ratio, so the [\ion{N}{2}]/[\ion{O}{2}] ratio is used to break the $R_{23}$ degeneracy.  If the measured line ratios correspond to an abundance outside of the range for which PT05 is calibrated ($6.8<$log(O/H)$+12<9.1$), we do not record the measurement.

Fifth, we apply the nitrogen abundance calibration of \cite{PVT} hereafter referred to as `PVT,'' relying on the combination of $P$ and [\ion{O}{3}], [\ion{N}{2}], and [S II] (``ONS'') line ratios.  The PVT diagnostic is calibrated separately for each of three different [\ion{N}{2}] regimes.  If the measured line ratios correspond to an abundance outside of the range for which PVT is calibrated ($7.3<$log(N/H)$+12<8.9$), we do not record the measurement.

We have not attempted to factor in the systematic error in the abundance diagnostics, although they are typically much larger ($\sim0.1$ dex) than our reported errors, which are derived by propagation of the line flux uncertainties.   For example, \cite{KE08} estimates the rms scatter between relative metallicities measured with the Z94 diagnostic, as compared to other popular diagnostics, is 0.07~dex based on a sample of 30,000 SDSS galaxies.  They find that the scatter in the other diagnostics are similar, the largest mean rms belonging to P01 (related to PT05) at 0.11~dex.  Because the only references for the accuracy of each abundance estimation technique are estimates from other diagnostics, which are not necessarily independent, any quantification of uncertainty must be interpreted with caution.

We have compared the oxygen abundance measurements made for the same HII region in different diagnostics.  Among the strong line methods, there is very good agreement between the Z94 and KD02 methods (standard deviation of \KDohtwoVSZninefourstd{} dex).  There is fairly good agreement between Z94 and N06 N2 (median offset of \NohsixNtwoVSZninefourmed{} dex and standard deviation of \NohsixNtwoVSZninefourstd{} dex) and between the N06 N2 and O3N2 diagnostics (negligible median offset, standard deviation of \NohsixOthreeNtwoVSNohsixNtwostd{} dex).  The PT05 diagnostic does not agree well with the other strong line methods, having a median offset as large as \PTohfiveVSZninefourmed{} dex (Z94) and a standard deviation as large as \PTohfiveVSNohsixNtwostd{} dex (N06 N2).

\begin{figure}
\centering
 	\plotone{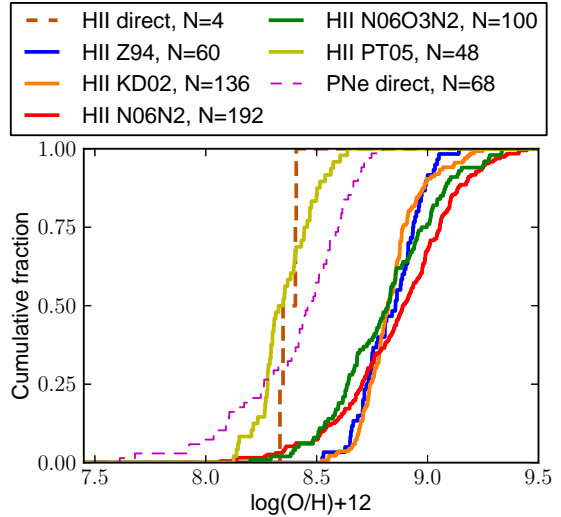}
\caption{The cumulative distribution functions (CDFs) of oxygen abundance for M31 HII regions and PNe (both disk and halo) as derived by different diagnostics.  The strong line methods are shown as solid lines, while the direct abundances are dashed.  The number (N) of objects for each diagnostic is noted in the legend and is limited by the available line flux ratios from each spectrum and the abundance range over which the diagnostic is calibrated.  PNe include halo objects.  A color version of this figure is available in the electronic journal.}\label{fig:abundCDF}
\end{figure}

In Figure \ref{fig:abundCDF} we show the cumulative distribution functions (CDFs) of oxygen abundance for the HII regions and PNe in our sample as derived by the different diagnostics.  The total range in PNe abundances is about $7.6\lesssim\log(\rm O/H)+12\lesssim8.8$ (a factor of 16).  The range of HII region abundances varies widely by diagnostic.  For example, PT05 abundances range from $8.0\lesssim\log(\rm O/H)+12\lesssim8.5$ while N06 N2 abundances range from $7.9\lesssim\log(\rm O/H)+12\lesssim9.5$.

The difference in both the shape and median value of the CDFs of different diagnostics are due to two factors: the systematic discrepancy between the calibrations, and the selection effects imposed by the requirement for certain emission lines to be detected in order to apply each diagnostic.    One additional selection effect is the range over which the diagnostics are calibrated; for example, the Z94 abundance scale is only calibrated down to log(O/H)$+12=8.4$, as described above.  If the diagnostic transformations of \cite{KE08} are applied (along with the trivial transformation between PP04 N2 and N06 N2), the CDFs for the N06 N2, KD02, and Z94 diagnostics agree reasonably well (Kolmogorov-Smirnov $p$-value $\gtrsim0.01$), despite the selection effects.  The discrepancy is largest between the PT05 diagnostic and the other strong line methods.  The median abundance for the N06 N2 diagnostic is log(O/H)$+12=\HIINohsixNtwomedian{}$, while for PT05 it is \HIIPTohfivemedian{} dex.  While this offset is approximately equal to the offset determined for the SDSS galaxies with these diagnostics \citep[][and also considering the conversion from N06 N2 to PP04 N2 and PT05 to P01]{KE08}.  However, \cite{KE08} determined that no reliable transformation can be established between PT05 and the other diagnostics because the relation is highly non-linear and has high scatter ($\gtrsim0.1$ dex). From Figure \ref{fig:abundCDF} it is clear that the direct abundances may only be derived for the lowest-metallicity objects where the auroral line is accessible.  

The combination of selection effects and calibration discrepancies should be kept in mind as we investigate the oxygen abundance profile of M31.  Results should only be compared if they are quoted in the same diagnostic, due to the systematic deviations and scatter that exist between different diagnostics.  Moreover, if a diagnostic does not probe the full range of abundances in the population, then the measured abundance profile in the diagnostic should not be used to infer physical properties of the galaxy.  In particular, we suggest that care be used in interpreting the abundance profile derived from the direct diagnostic, because it imposes severe selection effects (the availability of the auroral line) that selectively exclude high-metallicity objects from samples, and the PT05 diagnostic, because it is produces abundances discrepant with other strong line diagnostics.  However, measurements from these diagnostics are still useful.  The direct abundance is the only diagnostic based on a measurement of the true electron temperature of the nebulae, and it provides a useful comparison to Galactic work \citep[e.g.][]{Shaver83}.  Moreover, the direct method must be applied to PNe, for which strong line methods are not available.  The PT05 diagnostic is the most modern calibration of the $P$ method, which typically reflects the oxygen abundance as measured by the direct method more closely than other strong line methods \citep[][as well as our results above]{Pilyugin01h}.

\section{DISCUSSION}
\label{sec:disc}

Our analyses are primarily concerned with looking for radial trends in the interstellar medium properties of M31, with the goal of identifying any information that describes the chemical evolution history of the galaxy.  For this purpose, we focus on objects in the disk of the galaxy.  While all HII regions studied in this survey are attributed to the disk, a large population ($N=\CountPNhalo{}$) of PNe appear in projection to be outside of the disk ($|Y| > 4$ kpc); this is the halo population discussed in Section~\ref{sec:obs-galdist}.  We exclude these disk PNe from our analysis, except where explicitly described.

We look for radial trends in the optical extinction, oxygen abundance, and nitrogen abundance of HII regions and PNe in M31.   Our initial analyses, fitting linear trends and looking for correlations, are summarized in Table \ref{tab:fitparam}. This table summarizes the significance of radial trends in two different ways:

\begin{description}
\item[Bootstrap] We fit a line by ordinary least squares to the radial distribution of the parameter, and then repeat many times with resampling.  We simultaneously resample from the set of all objects with measurements of that parameter (with replacement) and also from the probability distribution function of the derived parameter.  In this way, we can estimate the slope and intercept (extrapolated value at the center of M31) of the radial trend in a way that is not sensitive to outliers or objects with poor-quality spectra.
\item[Spearman] We report the Spearman rank correlation statistic $\rho$ and its $p$ value.  A $\rho$ value much less than zero indicates a strongly negative gradient.  A $p$ value much greater than zero would indicate that any apparent correlation with radius could be due merely to chance.
\end{description}

We discuss these gradient analyses in the following section.

\subsection{Extinction}
\label{sec:disc-extinction}

The estimated values of extinction in the visual band ($A_V$) as derived from the Balmer decrement are reported in Tables~\ref{tab:derHII} and \ref{tab:derPN} and vary from $0$ to nearly $5$~mag.  \cite{blair82} sampled HII regions with extinction up to $\sim2.3$ mag and \cite{Galarza99}  up to $\sim4$~mag.

\begin{figure}
\centering
 	\plotone{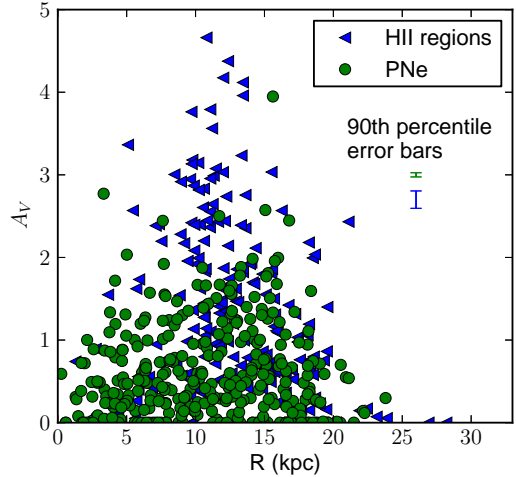}
\caption{Radial extinction profile for HII regions and disk PNe in M31.  The procedure for estimating $A_V$ is described in the text.  Typical error bars are smaller than the size of the points, but 90th percentile error bars are shown to illustrate the uncertainty for the least-constrained PNe (green) and HII regions (blue).}\label{fig:Av}
\end{figure}

Figure~\ref{fig:Av} illustrates the $A_V$ distribution versus galactocentric distance for \HIINAv\ HII regions and \PNNAv\ PNe.  The extinction is patchy.  Because reddening imposes a selection effect on our sample, the maximum extinction we observe at a given radius is just a lower limit.  Nonetheless, the maximum extinction varies radially: as high as $\sim5$~mag from $\sim10-15$ kpc, in the Ring of M31, to $\lesssim1.5$~mag beyond 20 kpc, in the outskirts of the disk.  However, objects fill the figure down to $\sim0$~mag at all radii.  The maximum extinction of PNe follow a similar radial trend, but typically have smaller extinction than HII regions ($A_{V, \rm PNe}\lesssim3$ mag).  It is expected that HII regions will have higher extinction than PNe, because they are near the large dust clouds associated with star formation and because they are primarily found in spiral arms rather than evenly throughout the disk.  PNe from the halo population are also likely to be found above the disk of M31 where extinction should be negligible.  Moreover, HII regions are typically an order of magnitude brighter than PNe and therefore may be observed through greater extinction \citep{panagia78}.  \cite{kumar79} asserted that a plateau at $\lesssim1$~mag in the extinction of HII regions begins at 12 kpc, while our data demonstrate that large ($>2$ mag)values of extinction are common out to nearly twice this distance.

PNe in the halo population have consistently small extinction values.  The median (and 16th, 84 percentile) value for extinction among halo PNe is $A_{V\rm halo}=\PNehaloAvFull$~mag for $N_{\rm halo}=\PNehalomedianAvN{}$, while for disk PNe it is much larger: $A_{V\rm disk}=\PNeAvFull$~mag for $N_{\rm disk}=\PNemedianAvN{}$.

\begin{figure}
\centering
        \plotone{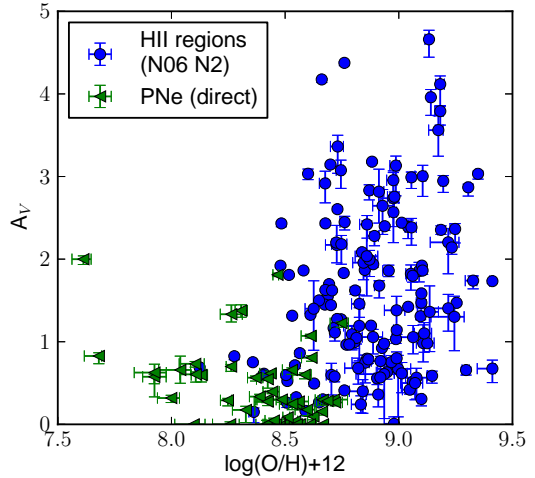}
\caption{The oxygen abundance of HII regions (N06~N2) and disk PNe (direct) in M31 vs the extinction ($A_V$) as measured from the Balmer decrement.  Statistical error bars (derived as described in the text) are shown for each dimension, but are typically too small to be visible.}\label{fig:abundVSAv}
\end{figure}

It is reasonable to expect the extinction to trace abundance, because dust grains that cause reddening are composed of heavy elements \citep{shields90}.  Some studies have reported such parallel gradients in spiral galaxies \citep[][for M33, M101, and M51, respectively]{Sarazin76,Viallefond86,Hulst88}, while others have measured flat extinction profiles \citep[][for M33 and M81]{Viallefond86,Kaufman87}.  In Figure~\ref{fig:abundVSAv} we show the oxygen abundance of HII regions and PNe in M31 against the extinction ($A_V$).  This figure illustrates that there is not a clear correlation between extinction and oxygen abundance among the objects in our survey.

Regardless of any trends in the extinction with radius or abundance, it is clear that the extinction in M31 is patchy.  For example, neighboring HII regions ($<0.5$ kpc in deprojected distance, or $\approx2.2$\arcmin\ separation on the sky) differ in extinction by as much as \HIIlocalAvdiff{} mag.  Among the \HIIlocalAvN{} such neighboring pairs in our sample with extinction measurements, 33\% have a discrepancy in $A_V$ of more than \HIIlocalAvtopthirdf{}~mag (Figure~\ref{fig:localdist}).  Some of this deviation is attributable to the large inclination of M31 introducing somewhat disparate column densities into the line of sight for apparently adjacent objects that are in front of and behind the disk.

\subsection{Radial oxygen abundance gradient}
\label{sec:disc-gradient}

The measurement of any abundance gradient among the M31 HII regions depends strongly on the choice of abundance diagnostic.  For example, if we employ the N06~N2 diagnostic, we measure a gradient (\HIIGradientNohsixNtwoBboot{}~dex~kpc$^{-1}$) that is negative at the $\sim4\sigma$-level and consistent with the canonical value ($-0.020 \pm 0.007$~dex  kpc$^{-1}$) of \cite{zaritsky94}.  If we instead use the Z94 diagnostic, we find a gradient that is much less steep and only different from zero at the $\sim1\sigma$ level (\HIIGradientZninefourBboot{}~dex~kpc$^{-1}$).  The N06 O3N2 (\HIIGradientNohsixOthreeNtwoBboot{}~dex~kpc$^{-1}$) and KD02 (\HIIGradientKDohtwoBboot{}~dex~kpc$^{-1}$) diagnostics yield similar results.  The $p$-value of the Spearman test suggests that a real correlation exists in all four cases, to varying degrees ($p_{\rm N2}=\HIISpearmanNohsixNtwop$, $p_{\rm Z94}=\HIISpearmanZninefourp$, $p_{\rm O3N2}=\HIISpearmanNohsixOthreeNtwop$, $p_{\rm KD02}=\HIISpearmanKDohtwop$).  If the PT05 diagnostic is used, however, we do not find a significant gradient (\HIIGradientPTohfiveBboot{}~dex~kpc$^{-1}$).  The Spearman test reflects the lesser significance of the radial trend in this diagnostics ($p_{\rm PT05}=\HIISpearmanPTohfivep$).  Moreover, different results can be achieved if the sample is divided by morphological type or surface brightness (Section~\ref{sec:disc-lumhii}).  We do not detect the temperature-sensitive auroral lines for enough HII regions to investigate the abundance gradient in the direct diagnostic.

\begin{figure*}
\centering
 	\plotone{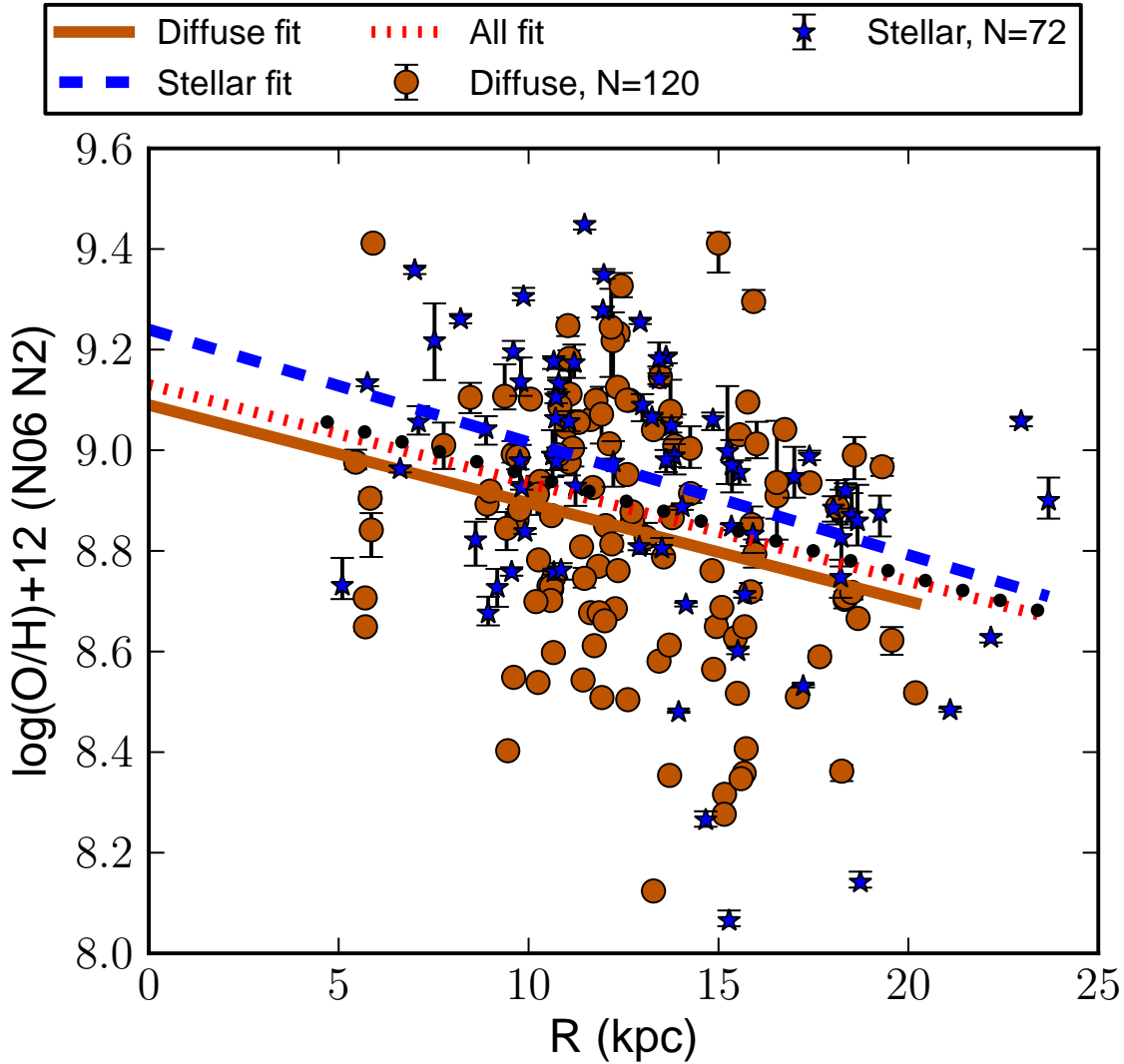}
\caption{The radial oxygen abundance profile of HII regions in M31 using the N06~N2 diagnostic.  Each point is color-coded by the surface brightness of the H$\alpha$ emission line, as discussed in Section~\ref{sec:obs-lineflux}.  The thick line is the best-fit linear abundance gradient, as derived using the bootstrap method described in the text.  The lower and upper dashed lines are constructed from the 33rd and 66th percentile slope and offset parameters calculated in the bootstrap simulation, respectively, shown to illustrate the possible variation in the fitted gradient.  The dotted line is the canonical gradient from \cite{zaritsky94}.  Statistical error bars (derived as described in the text) are shown for the oxygen abundance, but are frequently too small to be visible.}\label{fig:HIIprofile}
\end{figure*}

An illustrative radial oxygen abundance profile of M31 HII regions is shown in Figure~\ref{fig:HIIprofile}, using the N06~N2 diagnostic.  This diagnostic is highlighted because it relies on only the brightest emission lines and is insensitive to flux calibration and reddening correction.  It therefore produces reliable abundance estimates for a very large number of HII regions ($N=\HIINNohsixNtwo{}$).  Also shown is the abundance gradient as fit by the bootstrap method: (\HIIGradientNohsixOthreeNtwoAboot) dex + (\HIIGradientNohsixNtwoBboot) dex kpc$^{-1}$.    This represents a relatively shalow gradient among the nearby spiral galaxies studied by \cite{zaritsky94}, falling in the $\ZKHgradpKPC$ percentile range based on the $1\sigma$ errorbars quoted above.  Employing the isophotal radius $\rho_0=16$~kpc they define for M31, the size normalized gradient falls in the $\ZKHgradpRHO$ percentile range.

\begin{figure}
\centering
        \plotone{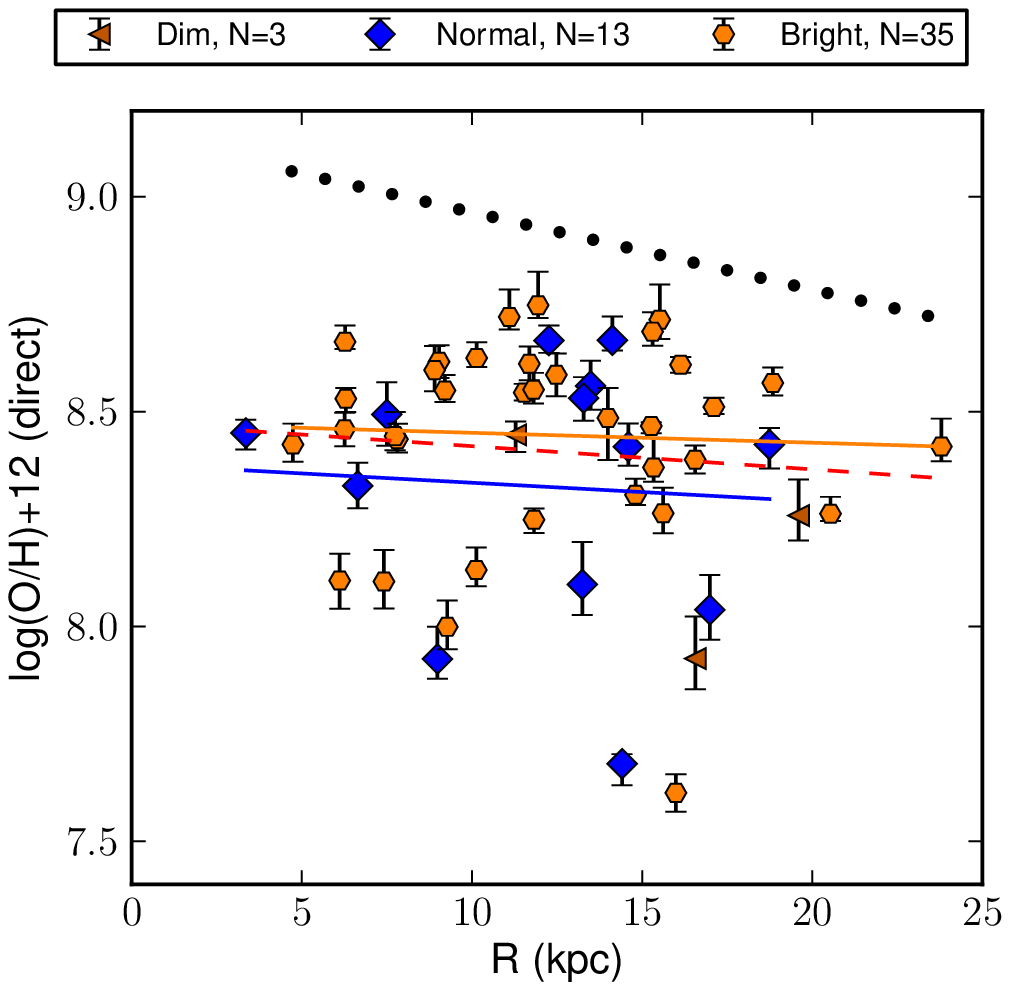}
\caption{The radial oxygen abundance profile of PNe in M31 using the direct diagnostic.  The sample of PNe is divided into 3 luminosity classes as described in  Section~\ref{sec:obs-lineflux}.  A radial gradient is fit to the sample of each flux class, and the dashed red line illustrates the gradient fit to the full sample.}\label{fig:PNprofile}
\end{figure}

The radial oxygen abundance profile of M31 PNe is shown in Figure~\ref{fig:PNprofile}, using the direct abundance diagnostic.  Also shown in the figure is the abundance gradient as fit by the bootstrap method, (\PNGradientdirectBboot) dex kpc$^{-1}$, which is consistent with zero.  The Spearman $p$-value is fairly large (\PNSpearmandirectp{}), emphasizing that there is no significant correlation between the PNe abundances and galactocentric radius.  However, various systematic effects influence the interpretation of this correlation, as we will discuss in Section~\ref{sec:disc-time}.

\subsection{Intrinsic scatter}
\label{sec:disc-scatter}

It is clear from Figures \ref{fig:HIIprofile} and \ref{fig:PNprofile} that, regardless of what the true slope of the abundance gradients may be, there is significant intrinsic scatter about the trend.  

\begin{figure}
\centering
        \plotone{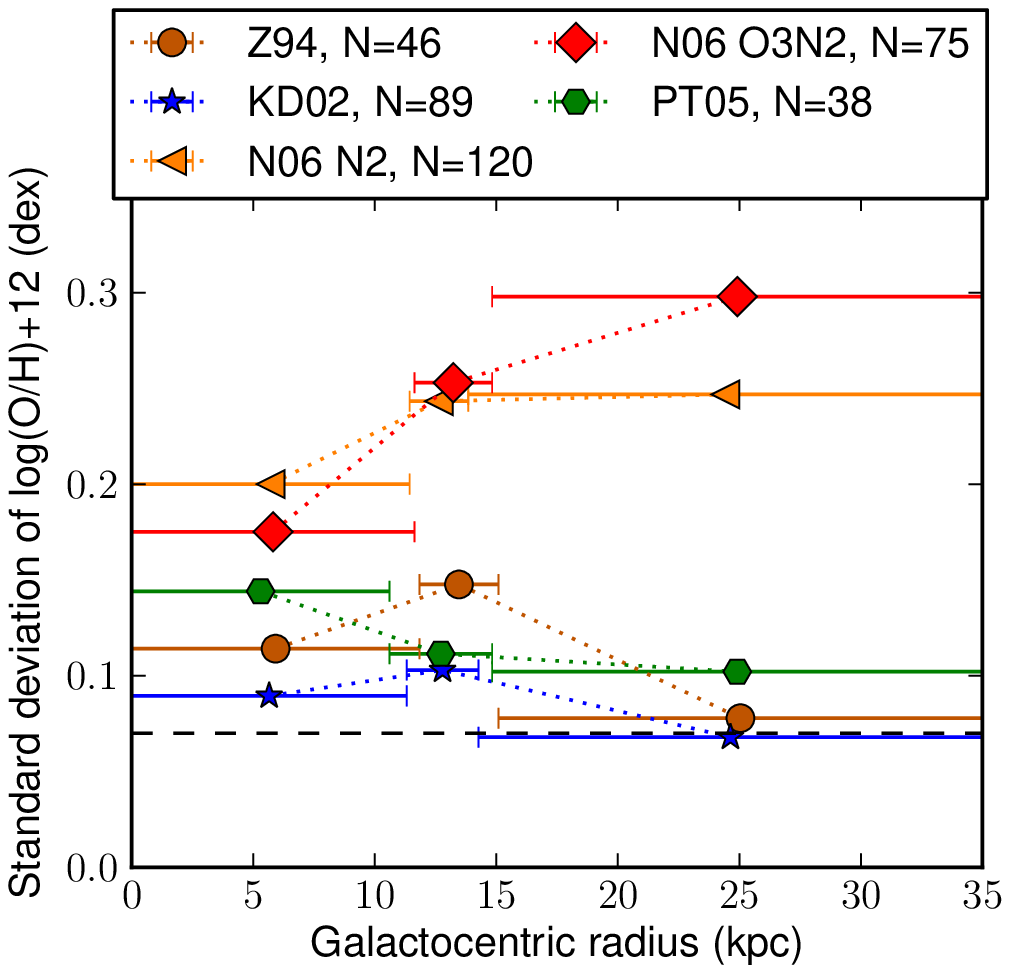}
\caption{The scatter in the radial oxygen abundance profile of the diffuse HII regions in M31.  The horizontal dashed line at 0.07~dex represents the typical systematic error in the diagnostic \citep[the mean rms error in the Z94 diagnostic from][]{KE08}.  The scatter is the standard deviation of the abundance of all HII regions in each of three different radial bins.  The bins are chosen such that each bin has an equal number of objects.  The bin edges are therefore different for each diagnostic, because the sample changes when different line ratios are required.  Various abundance diagnostics are employed, with sample sizes as noted in the legend.}\label{fig:intscat}
\end{figure}

In Figure~\ref{fig:intscat} we characterize this scatter by calculating the standard deviation of the abundance of all HII regions in different radial bins for each abundance diagnostic.  By dividing the sample into radial bins, as opposed to calculating the standard deviation of the entire sample, we partially remove the variance that would be introduced by an abundance gradient.  While the KD02, PT05, and Z94 diagnostics seem to produce the least scatter ($\sim0.1$~dex), this could be in part due to selection effects; the emission lines necessary for calculating the R$_{23}$ ratio are not accessible in fainter HII regions.  If we consider the diagnostics with the largest sample size, N06~N2, the scatter in abundance rises from $\sim0.2-0.3$~dex from the inner to outer regions of the disk.  This is significantly larger than the scatter inherent to the abundance diagnostics themselves \citep[e.g. $\sim 0.07$~dex for Z94 and PP04 N2,][]{KE08}.

The conclusion that the intrinsic scatter in the abundance gradient is larger than the observational uncertainty reflects some studies in the literature.  If the best-fit gradient is subtracted from the M31 HII region abundances measured in \cite{zaritsky94}, then the standard deviation among the abundances is $\sim0.16$~dex --- similar to what we measure with the Z94 diagnostic.  \cite{Rosolowsky08} found an intrinsic scatter of $0.11$~dex among their 61 HII regions in M33, and asserted that this is larger than the precision of the measurements.  However, \cite{Bresolin11} have argued that certain systematic effects have artificially increased the scatter measured by \cite{Rosolowsky08}, particularly the inclusion of high-excitation HII regions and low S/N spectra. Because only \HIINdirect\ of our spectra meet the strict S/N threshold suggested by \cite{Bresolin11} (e.g. S/N~([\ion{O}{3}]~$\lambda4363)~> 5$), our dataset is not sufficient to address the intrinsic scatter in a subset of the data as they recommend.  However, any high-excitation objects in our sample of the type discussed by \cite{Bresolin11} would instead be classified as PNe (Section~\ref{sec:obs-class}), and therefore would not contaminate the HII region statistics.  Moreover, the uncertainties in the strong line abundance measurements are negligible compared to the measures intrinsic scatter.  For example, for the N06~N2 diagnostic our measurement uncertainties as propagated from the emission line flux uncertainties have a median of \HIINohsixerrMedian~dex and 90th percentile value \HIINohsixerrNinety~dex, much smaller than the $\gtrsim0.2$~dex intrinsic scatter we measure in the radial abundance profile.  Future spectroscopic studies of HII regions in M31 should acquire spectra of sufficient S/N in [\ion{O}{3}])~$\lambda4363$ to measure the intrinsic scatter in the direct diagnostics.

\begin{figure*}
\centering
        \plotone{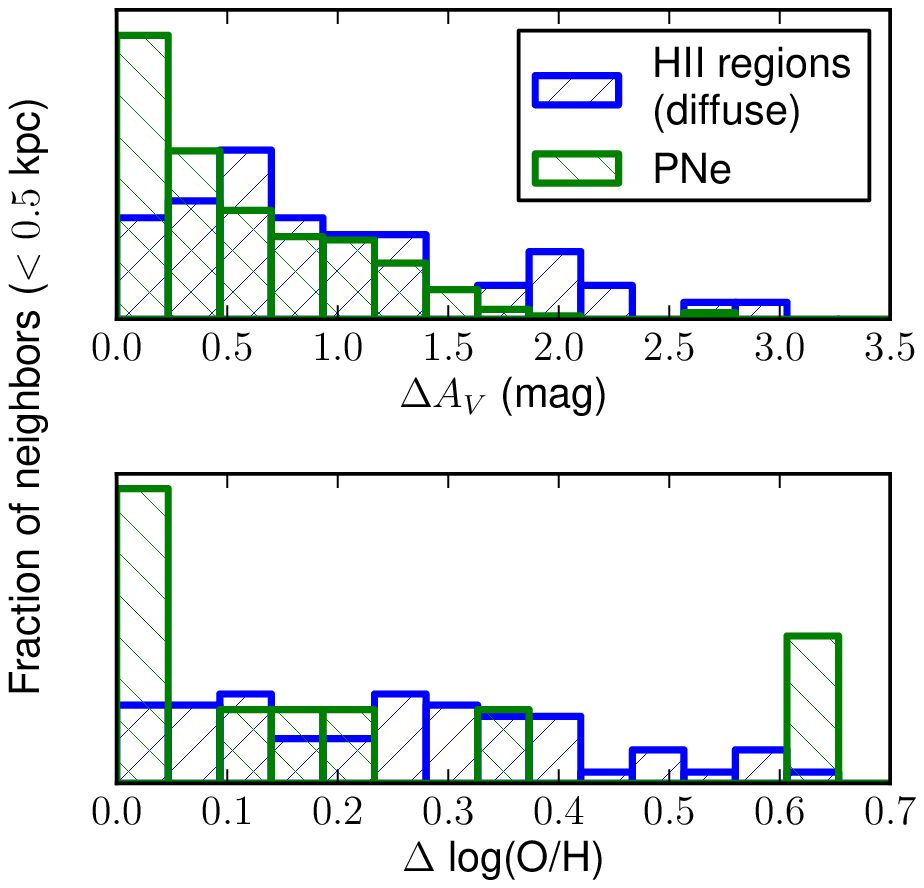}
\caption{The distribution of discrepancies between the extinction value and oxygen abundance (N06~N2 for HII regions, direct for PNe) of neighboring HII regions and PNe.  Neighbors are defined as any pair of objects that are separated by $<2.2$\arcmin\ on the sky, corresponding to $<0.5$~kpc in deprojected distance in M31.  Halo PNe are excluded.  There are \HIIlocalAvN{} such neighboring HII region pairs with extinction measurements and \HIIlocalOHN{} such pairs with oxygen abundance measurements in our sample; \PNlocalAvN{} and \PNlocalOHN{} PNe pairs, respectively.}\label{fig:localdist}
\end{figure*}

In Figure~\ref{fig:localdist}, we investigate local fluctuations in the ISM of M31.  We do so by considering the discrepancy in extinction and abundance measurements among HII regions and disk PNe separated by less than $\approx2.2$\arcmin\ on the sky, corresponding to $<0.5$ kpc in deprojected distance in M31.  As we have previously discussed in Section~\ref{sec:disc-extinction}, local fluctuations in extinction are often quite large --- $\Delta$(O/H)$>\HIIlocalOHtopthirdf{}$~dex and $\Delta A_V>\HIIlocalAvtopthirdf{}$~mag for one third of HII regions and $\Delta$(O/H)$>\PNlocalOHtopthirdf{}$~dex for one third of disk PNe.

As illustrated in Figure~\ref{fig:bothfinder}, the oxygen abundance of the ISM of M31 is inhomogeneous.  Neighboring HII regions ($<0.5$ kpc in deprojected distance) differ in oxygen abundance (N06~N2) by as much as \HIIlocalOHdiff{} dex, an order of magnitude.  Among the \HIIlocalOHN{} such neighboring pairs in our sample with N06~N2 abundance measurements, 33\% have a discrepancy in log(O/H) of more than \HIIlocalOHtopthirdf{} dex (Figure~\ref{fig:localdist}).  These discrepancies could be partially explained by measurement uncertainty, however $0.4$~dex is $\sim5\times$ the scatter expected from the systematic uncertainties in the diagnostic \citep[$\sim0.07$~dex for the similar PP04 N2 diagnostic,][]{KE08}.  This scatter is similar to the maximum discrepancy among the 8 PNe and 5 HII regions ($\sim0.3$ and $\sim0.2$~dex, respectively) in the immediate solar neighborhood ($\lesssim2$ kpc) observed by \cite{Rodriguez11}.  Only \PNlocalOHN{} neighboring PNe pairs have direct abundance estimates, so we do not consider their distribution of discrepancies.

We present spectra for an example of two neighboring HII regions with discrepant abundances in Figure~\ref{fig:spec}.  These are two diffuse HII regions (objects \IDnewpneightfournine{} and \IDnewpneighteightnine{}) that are separated by only \favsepMIN{}\arcmin\ on the sky, corresponding to a separation of $\sim\favsepKPC{}$ kpc at the distance of M31.  When we calculate their galactocentric radii, the difference is \favsepR{} kpc, and their velocities only differ by \favsepVEL{} km s$^{-1}$.  Despite being so nearby, object \IDnewpneightfournine{} is low metallicity (log(O/H)$_{\rm N06~N2}+12=\newpneightfournineNohsixNtwo{}$) and object \IDnewpneighteightnine{} is high metallicity (log(O/H)$_{\rm N06~N2}+12=\newpneighteightnineNohsixNtwo{}$).

While it would be interesting to search for non-linearity in the abundance profile (such as breaks near the well-known star-forming ring of M31), the high-level of intrinsic abundance scatter present throughout the disc would make it difficult to evaluate different models.

For PNe, we estimate the intrinsic scatter in the O abundance of M31 disk PNe as $\gtrsim\PNescatter{}$~dex.  Because we find no evidence of a significant radial trend in oxygen abundance for PNe, we simply calculate this number as the standard deviation of the \PNescatterN{} disk PNe with directly measured abundances.  We consider this to be a lower limit because high-metallicity PNe are systematically excluded from our direct abundance sample due to the weakness of the auroral line.  The median oxygen abundance is log(O/H)$+12=\PNemedian{}$~dex.  For the \PNehaloscatterN{} PNe in the halo of M31 for which we can measure direct abundances, we find a median value and standard deviation of log(O/H)$+12=\PNehalomedian{}\pm\PNehaloscatter{}$~dex.  The median abundance in the halo is therefore lower than that in the disk, but this distinction is small given the intrinsic scatter in the abundances of each population.  

\cite{Henry10} derive the oxygen abundance gradient in the Galaxy from observations of 124 PNe with high-quality spectra and well-determined distances.  They report a best-fit gradient of $-0.058\pm0.006$~dex kpc$^{-1}$.  They assert that the scatter around this best-fit gradient is $\sim40$\% larger than the uncertainties they ascribe to the abundance estimates.  Similarly, we find an intrinsic scatter in PNe abundances that is larger than the measurement error.  the scatter in the M31 disk PNe abundances we report above is $\gtrsim2\times$ larger than the median uncertainty in our disk PNe direct abundance estimates ($\sigma\sim\PNemedianerr{}$~dex, as derived by propagation of the line flux uncertainties).  An accounting of systematic errors could inflate the asserted measurement error, but they could not account for the intrinsic scatter unless they alter the measured abundances by a factor of $\gtrsim2$.

\cite{Rosolowsky08} and \cite{Magrini10} invoke inefficient azimuthal mixing to explain local fluctuations in the ISM metallicity of M33 (but see also \citealt{Bresolin11}).  In this scenario, mixing performed by velocity shear due to differential rotation occurs on a longer timescale ($\sim10^8$ yrs) than enrichment by star formation in the spiral arms of the galaxy.  Such a scenario could also apply to the inhomogeneities we observe in the ISM of M31.

\subsection{Dependence on HII region properties}
\label{sec:disc-lumhii}

Due to its size and proximity, M31 provides a unique laboratory for studying the ISM of a spiral galaxy; for many extragalactic studies, only the brightest HII regions in the galaxy are accessible to spectroscopy.  While the large number of relatively-dim HII regions included in our survey (Section~\ref{sec:obs-lineflux}) allows us to probe the ISM properties of M31 more thoroughly than ever before, it also has the potential to introduce discrepancies with past work.  Here we investigate whether the measured abundance profile varies systematically with the brightness or compactness of the HII regions in the sample.

\begin{figure}
\centering
        \plotone{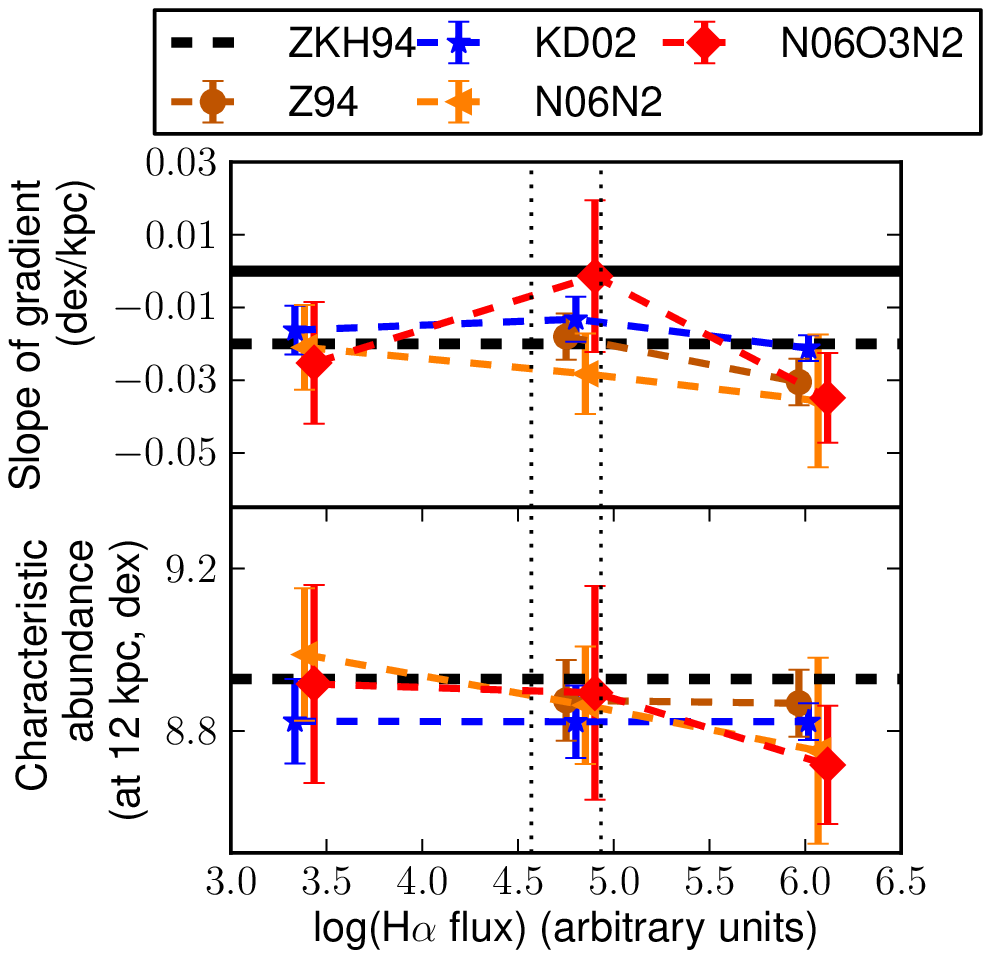}
\caption{The fitted parameters of the oxygen abundance (log(O/H)+12) gradient for diffuse HII regions in M31 in bins of H$\alpha$ emission line flux density.  The bins were chosen such that they each have an equal sample size, as described in Section~\ref{sec:obs-lineflux}.  The dotted vertical lines denote the width of the bin.  The points are given small x-offsets for clarity.  The y-errorbars are the bootstrap parameter uncertainties as described in the text.  Different abundance diagnostics are used as noted in the legend.}\label{fig:HIIslopeVSlum}
\end{figure}

In Figure~\ref{fig:HIIslopeVSlum} we investigate the fitted abundance gradient parameters (slope and characteristic abundance at 12~kpc) for the diffuse HII regions as a function of H$\alpha$ emission line flux density by dividing our sample into surface brightness bins (see Section~\ref{sec:obs-lineflux}).  Although the bins are comprised of equal numbers of HII regions, because the S/N of spectral lines depends strongly on surface brightness, there are typically fewer abundance measurements available in the lower surface brightness bins.  We adopt a minimum of 5 abundance measurements for performing abundance gradient analysis, which is the minimum number of HII regions sampled for any galaxy by \cite{zaritsky94}.  There are not sufficient PT05 abundance measurements to perform this analysis with that diagnostic.  We find that the fitted slope and characteristic abundance parameters are essentially independent of surface brightness, varying by amounts consistent with the error bars on the fitted parameters.

This result contrasts with that reported by \cite{Magrini10}, who found that the abundance gradient in M33 was more than twice as steep for bright ``giant'' HII regions than for the less-luminous objects in their sample.  Given the distance of M33 (840~kpc), their threshold for bright HII regions ($F_{\rm{H}\alpha}\approx1.2\times10^{34}$~ergs~s$^{-1}$) is similar to our threshold for normal/bright HII regions.  They attribute the surface-brightness dependence of the gradient to self-enrichment in giant HII regions.  However, \cite{Magrini10} derive metallicity gradients based on direct abundance measurements.  Because \cite{Magrini10} exclude $\sim1/3$ of their spectroscopic sample due to insufficient S/N in the emission lines necessary to derive direct abundances, one possible systematic explanation for the stronger gradient they measure among brighter HII regions is that the [\ion{O}{3}]~$\lambda4363$ line is weaker in high-metallicity HII regions.  Therefore high-metallicity HII regions may be systematically excluded from their lower-luminosity (not giant) HII region sample.  Similarly, the strong-line methods we apply in this study carry certain selection effects (see Section~\ref{sec:obs-abundance}), but we do not find that the gradient parameters vary significantly with brightness.

We can also investigate the abundance gradient as a function of HII region morphology (as described in Section~\ref{sec:obs-class}).  As described in Section~\ref{sec:obs-lineflux}, the morphological and surface brightness classifications are quite different, and the stellar HII regions may have PN contamination.  \cite{Galarza99} only found a significant radial abundance trend in the R$_{23}$ line ratio for "center-brightened" (stellar) HII regions, and no gradient for HII regions of more extended morphology (diffuse).  Generally, the abundance gradient slopes for diffuse HII regions reported in Table \ref{tab:fitparam} are in good agreement with the canonical value from \cite{zaritsky94}, while the slopes for stellar HII regions are more shallow or consistent with zero.  For example, in the Z94 diagnostic, the Spearman test indicates a significant gradient among diffuse HII regions ($p=\HIIdiffuseSpearmanZninefourp$), but not stellar ($p=\HIIstellarSpearmanZninefourp$).  In detail, Table \ref{tab:fitparam} demonstrate that for most abundance diagnostics (e.g. N06~N2), the radial gradient slopes derived from objects of either morphology agree to within their errorbars ($\Delta$log(O/H)$+12_{{\rm N06~N2~[diffuse,stellar]}}=$~[\HIIdiffuseGradientNohsixNtwoBboot{},~\HIIstellarGradientNohsixNtwoBboot{}]~dex~kpc$^{-1}$).  \cite{Galarza99} do not report a trend in [\ion{N}{2}]/H$\alpha$ for objects of any classification, whereas we report significant and similar abundance gradients for this line ratio in both diffuse ($\Delta$log([\ion{N}{2}]/H$\alpha$)$_{{\rm diffuse}}=$\HIIdiffuseGradientNtwoBboot{} dex kpc$^{-1}$) and stellar ($\Delta$log([\ion{N}{2}]/H$\alpha$)$_{{\rm stellar}}=$\HIIstellarGradientNtwoBboot{} dex kpc$^{-1}$) HII regions, equivalent to the gradient measured in N06~N2 oxygen abundance.  We do find a systematic offset in the N06N2 diagnostic, such that the best fit gradient intercept (metallicity at the galactic center) for the [diffuse,stellar] HII regions is $\rm{log(O/H)}+12=$[\HIIdiffuseGradientNohsixNtwoAboot, \HIIstellarGradientNohsixNtwoAboot] in the N06N2 diagnostic.  This can be explained by a systematic difference in the hardness of the ionizing radiation in HII regions of these two morphological classes, which causes strong line methods which do not account for the ionization parameter to overestimate the metallicity of compact HII regions.

However, there are particularly strong disagreements between stellar and diffuse HII regions among the KD02 and PT05 diagnostics (Table \ref{tab:fitparam}).  In detail, these discrepancies are driven by the presence of anomalously high-metallicity (log(O/H)~$+12>9.0$ on the KD02 scale) stellar HII regions.  Because these HII regions are also dim, with surface brightness consistent with PNe in our sample, they are likely to be PN contaminants.

\subsection{Time-variation in the abundance gradient}
\label{sec:disc-time}

The HII regions in our sample are in general more enriched than the PNe, as is demonstrated by Figure~\ref{fig:abundCDF}.  Similarly, the difference in the abundances of HII regions and PNe in M33 reported by \cite{Magrini10} was $\sim0.1$~dex and interpreted in the context of the time-varying composition of the ISM.  The median oxygen abundance and standard deviation for M31 disk PNe is log(O/H)$+12=\PNemedian{}\pm\PNescatter{}$~dex.  Among HII regions studied with the N06~N2 diagnostic, the typical abundances are much larger (log(O/H)$+12=\HIImedNohsixNtwo$).  This would suggest a similar discrepancy between HII regions and PNe as in M33; however, the median abundance we measure for PNe may be depressed because of the selection effect on the [\ion{O}{3}]~$\lambda4363$ line required to estimate direct abundances.  Moreover, using a diagnostic that typically correlates between with direct abundance measurements, PT05, we find a smaller median metallicity for HII regions (log(O/H)$+12=\HIImedPTohfive$); again, selection effects should act to exclude higher-metallicity objects.  This illustrates the complicating role of systematic effects in comparing abundance measurements for a statistical sample of extragalactic HII regions and PNe.

\begin{figure}
\centering
        \plotone{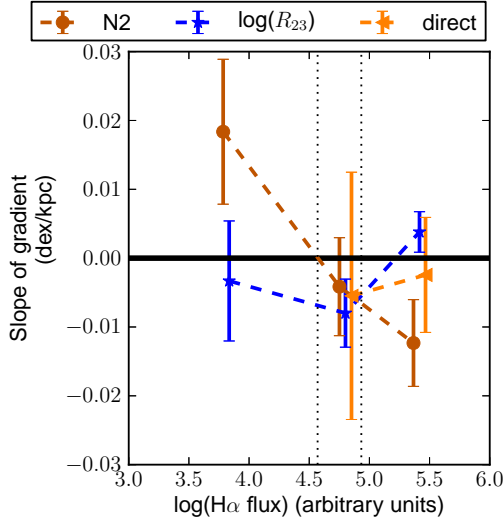}
\caption{The fitted parameters of the oxygen abundance (log(O/H)+12) gradient and line ratios for PNe in M31 in bins of H$\alpha$ emission line flux density.  The figure is constructed similarity to Figure~\ref{fig:HIIslopeVSlum}.}\label{fig:PNslopeVSlum}
\end{figure}

In Figure~\ref{fig:PNslopeVSlum}, we bin the PNe by surface brightness to investigate the potential time-dependence of the abundance gradient.  As discussed in Section~\ref{sec:intro}, the luminosity of PNe are related to the masses of their progenitor stars and therefore to their ages. As in Section~\ref{sec:disc-lumhii}, we adopt a minimum of 5 measurements for performing gradient analysis in each bin, and we are therefore not able to fit an abundance gradient for the least bright PNe.

If the radial abundance gradient has strengthened over time, we would expect to find a gradient that is more negative with increasing PN surface brightness.  In fact, the trend in the direct abundance is never strongly inconsistent with zero for either surface brightness bin.  Among the brightest PNe, the slope is \PNBrightBbootdirect~dex~kpc$^{-1}$.  We do find that the median metallicity increases with brightness class, with [$\PNdirectmedLone$,$\PNdirectmedLtwo$,$\PNdirectmedLthree$]~dex for [Dim,~Normal,~Bright] PNe.  However, because the auroral lines are only detectable in lower-metallicity PNe, selection effects could be eliminating high-metallicity PNe from our sample which could reveal a significant abundance gradient.  Moreover, this selection effect would act more strongly to remove high-metallicity objects among the dimmer PNe, which would mimic the signature of increasing metallicity with PN surface brightness.

Because direct abundance estimates can only be made for \PNNdirect{} of our \CountPNdisk{} disk PNe spectra, it is worthwhile to look for gradients in strong line ratios such as $R_{23}$ and [\ion{N}{2}]/H$\alpha$.  While these PN line ratios are not directly abundance-sensitive as they are in HII regions, significant gradients (if present) could indicate interesting trends in other physical properties.  
For [\ion{N}{2}]/H$\alpha$, the gradient is positive at the $\sim2\sigma$ level for the lowest-surface brightness bin (\PNDimBbootNtwo~dex~kpc$^{-1}$), and increases significantly with surface brightness such that it is positive among the brightest PNe (\PNBrightBbootNtwo~dex~kpc$^{-1}$).  The existence of this N2 gradient, and its correlation with surface brightness, could indicate a time-varying gradient in excitation, chemical composition, or both.  For R$_{23}$, the slope is never significantly different from zero in any brightness bin.

It would be difficult to distinguish time evolution in the abundance profile of M31 by comparing its PNe to HII regions, due to the intrinsic scatter in the populations and the uncertainties in the determination of the abundance gradients.  Results would be particularly influenced by the choice of strong line diagnostic for HII region abundances, and by the cuts made on morphology/surface brightness (Section~\ref{sec:disc-gradient}).

\subsection{HII region nitrogen abundance gradient}
\label{sec:disc-N}

While the oxygen abundance gradient is the most observationally accessible (Section~\ref{sec:intro}), the radial nitrogen gradient is of particular interest because it may be the steepest of any observable element \citep[e.g. in M33,][]{Magrini10}.  The models of \cite{Magrini10} show that the N gradient should be steeper than the O gradient due to the different timescales for production; N is produced primarily in low and intermediate-mass stars, while high-mass stars more efficiently produce O.

\begin{figure}
\centering
 	\plotone{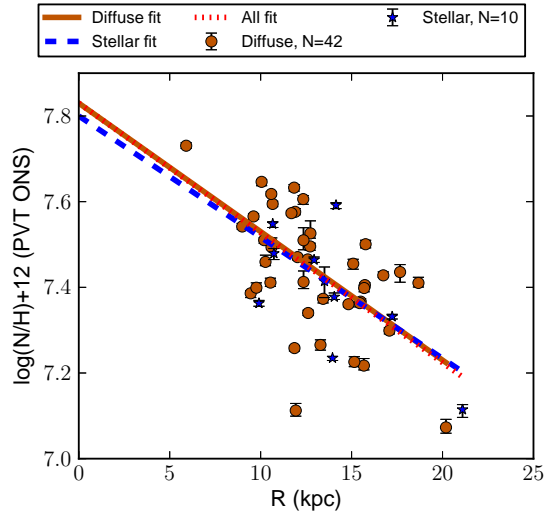}
\caption{The radial nitrogen abundance profile of HII regions in M31 using the diagnostic of \cite{PVT}.  The figure is constructed similarly to Figure~\ref{fig:HIIprofile}}\label{fig:HIInitrogenprofile}
\end{figure}

The radial nitrogen abundance profile of M31 HII regions is shown in Figure~\ref{fig:HIInitrogenprofile}, using the diagnostics of \cite{PVT}.  Also shown is the highly-significant ($\sim4\sigma$) radial gradient for the diffuse objects as fit by the bootstrap method, (\HIIdiffuseGradientPVTNHONSAboot{}) dex + (\HIIdiffuseGradientPVTNHONSBboot{}) dex kpc$^{-1}$.  This slope is approximately as steep or steeper than all the oxygen abundance gradients reported for any diagnostic and morphological selection in Table \ref{tab:fitparam}.  However, it is only one third as steep as the nitrogen gradient in M33 \citep[$-0.08\pm0.03$~dex kpc$^{-1}$,][]{Magrini10}.  If we include the stellar HII regions ($N=\HIIstellarNPVTNHONS$), we find a gradient that is similar (\HIIGradientPVTNHONSBboot{} dex kpc$^{-1}$).  

As for the oxygen abundances, we find a large intrinsic scatter about this gradient.  Subtracting the fitted trend among all the HII regions, we find an rms scatter of \NHrms~dex.  This is significantly larger ($\sim2$ times) the systematic uncertainty attributable to the strong line diagnostic of \citep{PVT}, who report an rms scatter of 0.05~dex (smaller than for the equivalent oxygen abundance diagnostic).

\subsection{Comparison to previous observations of M31}
\label{sec:disc-compare}

A variety of authors have previously derived abundance gradients for M31 from other surveys of HII regions, as well as O and B stars.

As we have discussed, \cite{dennefeld81}, \cite{blair82}, \cite{zaritsky94}, and \cite{Galarza99} have previously derived the abundance gradient of M31 from surveys of HII regions.  From those works, we have the canonical result of $-0.020 \pm 0.007$~dex kpc$^{-1}$, produced by \cite{zaritsky94} from the combined sample of 19 HII regions from \cite{dennefeld81} and \cite{blair82}.  In their survey of 46 HII regions in M31, \cite{Galarza99} found agreement with this gradient, but only among center-brightened HII regions.  Among HII regions of other morphologies, they found no significant trends.

\cite{Trundle02} derived oxygen abundances for 7 stars of type O and B from about 5 to 30 kpc in the disk of M31.  Additionally, they provided a re-analysis of the HII regions of \cite{blair82} using modern diagnostics, producing gradients that range from $-0.013$ (P-method) to $-0.027$ (M91) dex kpc$^{-1}$.  Their least squares fit to the stars yielded an abundance gradient of $-0.017\pm0.02$~dex kpc$^{-1}$, in good agreement with the canonical result.  However, they noted that if they omitted the possible multiple system OB 8-76, then they instead measured a slope consistent with zero ($-0.006\pm0.02$~dex kpc$^{-1}$).  \cite{worthey05} produced color-magnitude diagrams from HST images to derive the stellar abundance gradient in M31, finding (with some uncertainty in their zero point) good agreement between the upper bound of the metallicity of the stellar population and the HII regions surveyed by \cite{blair82} and \cite{dennefeld81} at a given radius.

The results of previous surveys of HII regions and high-mass stars suggest that any discrepancy between the M31 abundance gradients derived from different sources are dependent on systematic effects related to sample selection and diagnostic.  Similarly, \cite{Urbaneja05} have found that the nebular abundance gradient can be smaller than, similar to, or larger than the stellar abundance gradient in the late-type spiral NGC 300 depending on the choice of diagnostic.

\section{CONCLUSIONS}
\label{sec:conc}

We have presented optical spectroscopy of an unprecedented sample of HII regions and PNe in a massive spiral galaxy, M31.  In total, we reported line flux measurements for $\CountHiiTotal{}$ HII regions and $\CountPNTotal{}$ PNe.  We have derived the extinction, nitrogen abundance, and oxygen abundance for subsets of these objects using a variety of methods, as described in Section~\ref{sec:observations}.  From the analysis of these observations, we emphasize the following conclusions:

\begin{enumerate}
\item For HII regions, we find an oxygen abundance gradient generally consistent with that found by \cite{zaritsky94}.  Using the N06 N2 diagnostic, we find a gradient of (\HIIGradientNohsixNtwoBboot)~dex~kpc$^{-1}$ among \HIINNohsixOthreeNtwo HII regions.  We find a significantly steeper gradient in nitrogen abundance, (\HIIGradientPVTNHONSBboot)~dex~kpc$^{-1}$ among $\HIINPVTNHONS$~HII regions.  These represent relatively shallow gradients as compared to other nearby spiral galaxies.
\item For PNe, we detect no significant oxygen abundance gradient among \PNNdirect\ objects for which the [O III]~$\lambda4363$ line is detected and the direct method can be applied.  However, using the line ratio [N II]/H$\alpha$, which is measurable for most PNe ($N=\PNNNtwo{}$), we find significant gradients which vary systematically with PN brightness, as illustrated in Figure \ref{fig:PNslopeVSlum}.
\item The ISM of M31 is highly inhomogeneous.  Both the visual extinction ($A_V$) and oxygen abundance vary significantly among even very nearby HII regions (\HIIlocalAvdiff{} mag and \HIIlocalOHdiff{} dex for some HII regions separated by $<2$ kpc, see Figures \ref{fig:bothfinder} and \ref{fig:localdist}).  Moreover, the intrinsic scatter observed about the HII region oxygen abundance gradient ($\sim0.1-0.3$ dex, see Figure \ref{fig:intscat}) is larger than the uncertainty inherent to the strong line diagnostics ($\sim 0.1$ dex).  Similarly, the scatter among PNe in our sample is $\gtrsim\PNescatter{}$~dex.
\item The abundance gradient derived for HII regions in M31 is dependent upon the strong line metallicity diagnostic employed, and can be affected systematically by sample characteristics such as HII region morphology and surface brightness.  In particular, for observations to a given depth, some diagnostics can only be applied to low or high-metallicity HII regions, unless they have sufficient surface brightness.  Among more compact (not extended, i.e. ``stellar'') nebulae that are spectroscopically classified as HII regions, we find evidence of PN contamination that can lead to erroneous strong line abundance measurements.  Sample characteristics of this type should be taken into careful consideration to mitigate systematic effects in future surveys of the ISM of nearby spiral galaxies.
\end{enumerate}

\acknowledgements
\label{sec:ackn}

The authors wish to thank the anonymous referee for helpful comments and Jack Baldwin, Pauline Barmby, Richard Henry, Christine Jones, Emily Levesque, Marie Machacek, Phil Massey, John Raymond, Ricardo Schiavon, Evan Skillman, and Jay Strader for their insights.

This work was supported by the National Science Foundation through a Graduate Research Fellowship provided to NES.

This work was supported in part by the National Science Foundation Research Experiences for Undergraduates (REU) and Department of Defense Awards to Stimulate and Support Undergraduate Research Experiences (ASSURE) programs under Grant no. 0754568 and by the Smithsonian Institution.

\bibliographystyle{fapj.bst}

\clearpage
\begin{deluxetable}{p{25pt}rrrrrrrlll}
\tabletypesize{\scriptsize}
\tablecaption{Basic data for M31 HII regions and PNe\label{tab:obj}}
\tablehead{\colhead{ID} & \colhead{RA} & \colhead{DEC} & \colhead{R (kpc)\tablenotemark{a}} & \colhead{Morph. Type\tablenotemark{b}} & \colhead{SB\tablenotemark{c}} & \colhead{ADC\tablenotemark{d}} & \colhead{Velocity (km~s$^{-1}$)} & \colhead{M06\tablenotemark{e}} & \colhead{RBC\tablenotemark{f}}  & \colhead{AMB\tablenotemark{g}}}
\startdata 
\cutinhead{HII regions}
HII001 & 0:37:24.12 & +40:17:56.2 & 23.0 & s & 2 & n & -483.6 &  &  &  \\
HII002 & 0:37:29.91 & +40:15:37.2 & 21.9 & s & 1 & n & -517.5 &  &  &  \\
HII003 & 0:37:47.35 & +39:51:30.8 & 23.7 & s & 1 & n & -492.6 &  &  &  \\
HII004 & 0:37:59.17 & +40:15:37.2 & 19.2 & s & 1 & n & -476.2 & M2372 &  &  \\
HII005 & 0:38:22.51 & +40:10:52.8 & 18.4 & s & 2 & n & -564.5 &  &  &  \\
HII006 & 0:38:39.79 & +40:34:48.0 & 18.2 & d & 1 & y & -469.3 &  &  & HII9 \\
HII007 & 0:38:41.29 & +39:47:40.0 & 28.2 & s & 1 & y & -526.8 &  &  &  \\
HII008 & 0:39:03.75 & +39:53:28.9 & 26.9 & s & 1 & y & -479.0 &  &  &  \\
HII009 & 0:39:07.69 & +40:40:05.1 & 16.2 & s & 2 & y & -486.1 &  &  & HII1 \\
HII010 & 0:39:13.09 & +40:41:13.9 & 15.9 & s & 1 & y & -488.5 &  &  & HII1 \\
\nodata & & & & & & & & \\
\cutinhead{Planetary nebulae}
PN001 & 0:38:44.18 & +40:17:58.8 & 16.6 & \nodata & 1 & y & -465.4 & M2364 &  & PN7 \\
PN002 & 0:39:02.52 & +40:22:50.5 & 15.2 & \nodata & 3 & y & -529.4 & M2964 &  & PN9 \\
PN003 & 0:39:05.33 & +40:41:45.9 & 17.1 & \nodata & 2 & y & -387.9 & M3198 &  &  \\
PN004 & 0:39:06.59 & +40:14:59.7 & 17.1 & \nodata & 3 & y & -560.0 & M2357 &  &  \\
PN005 & 0:39:14.80 & +40:24:28.5 & 14.6 & \nodata & 1 & y & -338.0 & M2304 &  & PN1 \\
PN006 & 0:39:15.00 & +40:26:34.5 & 14.3 & \nodata & 3 & y & -538.8 & M2308 &  & PN1 \\
PN007 & 0:39:15.80 & +40:12:38.4 & 18.3 & \nodata & 2 & y & -488.8 & M2972 &  &  \\
PN008 & 0:39:16.29 & +40:22:12.8 & 15.0 & \nodata & 2 & y & -551.0 & M2960 &  & PN1 \\
PN009 & 0:39:18.51 & +40:09:18.9 & 19.9 & \nodata & 1 & y & -379.4 & M2404 &  &  \\
PN010 & 0:39:26.12 & +40:44:25.9 & 15.2 & \nodata & 1 & y & -407.7 & M1967 &  &  \\
\nodata & & & & & & & & \\
\cutinhead{Planetary nebulae (halo population)}
PNh001 & 0:35:50.74 & +42:21:04.4 & 104.8 & \nodata & 2 & y & -296.8 & M2543 &  &  \\
PNh002 & 0:36:27.15 & +42:06:21.8 & 89.9 & \nodata & 1 & y & -15.2 & M2549 &  &  \\
PNh003 & 0:37:09.28 & +42:38:18.9 & 103.6 & \nodata & 1 & y & -245.0 & M2542 &  &  \\
PNh004 & 0:37:28.34 & +42:10:57.6 & 83.3 & \nodata & 1 & y & -231.0 & M2544 &  &  \\
PNh005 & 0:37:54.37 & +42:14:49.3 & 81.7 & \nodata & 1 & y & -334.2 & M7 &  &  \\
PNh006 & 0:38:48.46 & +41:39:37.3 & 51.2 & \nodata & 2 & y & -195.4 & M352 &  &  \\
PNh007 & 0:38:56.63 & +39:47:14.1 & 29.7 & \nodata & 2 & y & -417.0 &  &  &  \\
PNh008 & 0:39:01.08 & +41:51:11.2 & 56.5 & \nodata & 3 & y & -216.0 & M174 &  &  \\
PNh009 & 0:39:21.16 & +42:25:16.3 & 75.2 & \nodata & 1 & y & -331.8 &  &  &  \\
PNh010 & 0:39:31.59 & +42:11:56.7 & 65.0 & \nodata & 1 & y & -343.5 & M5 &  &  \\
\nodata & & & & & & & & \\
\cutinhead{Unclassified}
X001 & 0:39:06.10 & +40:37:22.8 & 15.7 & s & 1 & y & -512.5 &  &  & HII1 \\
X002 & 0:39:07.30 & +40:36:25.5 & 15.4 & s & 1 & y & -513.9 &  &  & HII1 \\
X003 & 0:39:08.89 & +40:24:11.5 & 14.8 & s & 1 & y & -534.1 &  &  & HII1 \\
X004 & 0:39:09.40 & +40:29:16.0 & 14.4 & s & 1 & y & -542.2 &  &  & HII1 \\
X005 & 0:39:12.90 & +40:50:59.6 & 19.7 & s & 1 & y & -408.2 &  &  & HII1 \\
X006 & 0:39:16.09 & +40:43:55.5 & 16.4 & s & 1 & y & -530.5 &  &  & HII1 \\
X007 & 0:39:16.90 & +40:20:55.5 & 15.4 & s & 1 & n & -560.4 &  &  & HII1 \\
X008 & 0:39:24.30 & +40:48:07.5 & 16.8 & s & 1 & y & -469.8 &  &  & HII2 \\
X009 & 0:39:34.80 & +40:49:59.5 & 16.1 & s & 1 & y & -446.3 &  &  & HII2 \\
X010 & 0:40:10.29 & +40:45:19.0 & 10.4 & s & 1 & y & -533.0 &  &  & HII5 \\
\nodata & & & & & & & & \\
\enddata
\tablecomments{The table is divided into sections based on the spectroscopic classification of the object as either an HII region (HII), planetary nebula (PN), PN in the halo population (PNh), or unclassified (X).  This classification is based on line ratio diagnostics, as described in \S \ref{sec:obs-class}.  Table \ref{tab:obj} is published in its entirety in the electronic edition of the {\it Astrophysical Journal} and at \url{https://www.cfa.harvard.edu/~nsanders/papers/M31/summary.html}.  A portion is shown here for guidance regarding its form and content.}
\tablenotetext{a}{The galactocentric radius of the object, calculated as described in \S \ref{sec:obs-galdist}.}
\tablenotetext{b}{The morphological classification of the object based on LGGS H$\alpha$ imaging, as described in \S \ref{sec:obs-class}.}
\tablenotetext{c}{The surface brightness class of the object as defined in \S \ref{sec:obs-lineflux} based on the H$\alpha$ flux.}
\tablenotetext{d}{The status of the atmospheric dispersion compensator during the observation of this object: ``y'' indicates that it was functioning, ``n'' indicates that it was not.}
\tablenotetext{e}{The ID number of the object from the PNe catalog of \cite{merrett06}.}
\tablenotetext{f}{The name of the object from version 3.5 of the Revised Bologna Catalog, \cite{Bologna}.}
\tablenotetext{e}{The ID number of the object from the HII region catalog of \cite{Azimlu11}.}
\end{deluxetable}
\clearpage
\begin{deluxetable}{lp{46pt}p{46pt}p{46pt}p{46pt}p{46pt}p{46pt}p{46pt}p{46pt}p{46pt}}
\tabletypesize{\scriptsize}
\tablecaption{Measured line fluxes for M31 HII regions and PNe, relative to H$\beta$\label{tab:mea}}
\tablehead{\colhead{ID} & \colhead{[O II]} & \colhead{[O III]} & \colhead{[O III]} & \colhead{[O III]} & \colhead{[N II]} & \colhead{H$\alpha$} & \colhead{[N II]} & \colhead{[S II]} & \colhead{[S II]} \\ 
      & \colhead{$\lambda$ 3727} & \colhead{$\lambda$ 4363} & \colhead{$\lambda$ 4959} & \colhead{$\lambda$ 5007} & \colhead{$\lambda$ 6548} & \colhead{$\lambda$ 6562} & \colhead{$\lambda$ 6584} & \colhead{$\lambda$ 6717} & \colhead{$\lambda$ 6731}}
\startdata 
\cutinhead{HII regions}
HII001 & $50\pm20$ & \nodata & \nodata & \nodata & $48\pm3$ & $442\pm5$ & $150\pm3$ & $57\pm4$ & $45\pm5$ \\
HII002 & \nodata & \nodata & \nodata & \nodata & $54\pm6$ & $255\pm4$ & $151\pm5$ & \nodata & \nodata \\
HII003 & \nodata & \nodata & \nodata & \nodata & $26\pm6$ & $264\pm6$ & $76\pm7$ & $22\pm4$ & $20\pm4$ \\
HII004 & \nodata & \nodata & $40\pm10$ & $100\pm10$ & \nodata & $252\pm8$ & $70\pm8$ & \nodata & \nodata \\
HII005 & \nodata & \nodata & \nodata & \nodata & $87\pm9$ & $409\pm9$ & $270\pm10$ & \nodata & \nodata \\
HII006 & $430\pm40$ & \nodata & \nodata & $110\pm20$ & \nodata & $340\pm20$ & \nodata & \nodata & \nodata \\
HII007 & \nodata & \nodata & $299\pm6$ & $852\pm9$ & \nodata & $213\pm6$ & \nodata & \nodata & \nodata \\
HII008 & \nodata & $15\pm4$ & $286\pm5$ & $830\pm10$ & \nodata & $221\pm5$ & \nodata & \nodata & \nodata \\
HII009 & \nodata & \nodata & \nodata & $50\pm10$ & \nodata & $606\pm10$ & \nodata & \nodata & \nodata \\
HII010 & $270\pm10$ & \nodata & \nodata & \nodata & $27\pm6$ & $306\pm6$ & $82\pm6$ & $43\pm6$ & $33\pm6$ \\
\nodata & & & & & & & & & \\
\cutinhead{Planetary nebulae}
PN001 & \nodata & \nodata & $383\pm5$ & $1115\pm10$ & \nodata & $247\pm9$ & \nodata & \nodata & \nodata \\
PN002 & \nodata & \nodata & $462\pm5$ & $1360\pm10$ & $29\pm3$ & $411\pm5$ & $75\pm3$ & \nodata & \nodata \\
PN003 & \nodata & \nodata & $149\pm4$ & $442\pm4$ & \nodata & $312\pm5$ & $23\pm6$ & \nodata & \nodata \\
PN004 & $34\pm3$ & $13\pm1$ & $440\pm4$ & $1300\pm10$ & $16\pm2$ & $292\pm3$ & $47\pm2$ & \nodata & \nodata \\
PN005 & \nodata & \nodata & $470\pm10$ & $1410\pm10$ & \nodata & $320\pm10$ & \nodata & \nodata & \nodata \\
PN006 & $70\pm20$ & \nodata & $312\pm5$ & $930\pm10$ & $50\pm6$ & $443\pm6$ & $146\pm6$ & \nodata & \nodata \\
PN007 & $110\pm20$ & \nodata & $191\pm3$ & $562\pm6$ & $39\pm8$ & $322\pm8$ & $113\pm8$ & \nodata & \nodata \\
PN008 & $80\pm10$ & \nodata & $381\pm5$ & $1160\pm10$ & $27\pm7$ & $342\pm8$ & $73\pm8$ & $18\pm6$ & \nodata \\
PN009 & \nodata & \nodata & $31\pm5$ & $100\pm5$ & \nodata & \nodata & \nodata & \nodata & \nodata \\
PN010 & $100\pm20$ & \nodata & $220\pm10$ & $650\pm10$ & \nodata & $224\pm5$ & $27\pm6$ & \nodata & \nodata \\
\nodata & & & & & & & & &\\
\cutinhead{Planetary nebulae (halo population)}
PNh001 & \nodata & \nodata & $414\pm6$ & $1240\pm10$ & $9\pm3$ & $294\pm3$ & $25\pm3$ & \nodata & \nodata \\
PNh002 & \nodata & $18\pm5$ & $421\pm7$ & $1240\pm10$ & \nodata & $259\pm4$ & $22\pm5$ & \nodata & \nodata \\
PNh003 & \nodata & \nodata & $470\pm10$ & $1380\pm10$ & \nodata & $310\pm10$ & $40\pm10$ & \nodata & \nodata \\
PNh004 & \nodata & \nodata & $380\pm20$ & $1090\pm20$ & \nodata & $260\pm20$ & \nodata & \nodata & \nodata \\
PNh005 & \nodata & \nodata & $120\pm10$ & $360\pm20$ & \nodata & $100\pm30$ & \nodata & \nodata & \nodata \\
PNh006 & $90\pm20$ & \nodata & $360\pm5$ & $1062\pm9$ & \nodata & $291\pm6$ & $52\pm6$ & \nodata & $14\pm4$ \\
PNh007 & $107\pm6$ & \nodata & $120\pm2$ & $345\pm3$ & $59\pm4$ & $264\pm5$ & $164\pm5$ & \nodata & \nodata \\
PNh008 & $48\pm6$ & $5\pm1$ & $337\pm3$ & $989\pm9$ & $12.2\pm0.9$ & $284\pm2$ & $41\pm1$ & \nodata & \nodata \\
PNh009 & \nodata & \nodata & $130\pm10$ & $380\pm10$ & \nodata & $100\pm20$ & \nodata & \nodata & \nodata \\
PNh010 & \nodata & \nodata & $140\pm20$ & $420\pm10$ & \nodata & $100\pm20$ & \nodata & \nodata & \nodata \\
\nodata & & & & & & & & & \\
\cutinhead{Unclassified}
X001 & \nodata & \nodata & \nodata & \nodata & \nodata & $100\pm6$ & $28\pm6$ & $20\pm6$ & \nodata \\
X002 & $60\pm10$ & \nodata & \nodata & \nodata & \nodata & $100\pm10$ & \nodata & $22\pm7$ & \nodata \\
X003 & \nodata & \nodata & \nodata & \nodata & \nodata & $100\pm20$ & \nodata & \nodata & \nodata \\
X004 & \nodata & \nodata & \nodata & \nodata & \nodata & $100\pm10$ & \nodata & \nodata & \nodata \\
X005 & \nodata & \nodata & \nodata & \nodata & \nodata & $100\pm7$ & $25\pm8$ & \nodata & \nodata \\
X006 & \nodata & \nodata & \nodata & \nodata & \nodata & $100\pm10$ & $70\pm10$ & \nodata & \nodata \\
X007 & \nodata & \nodata & \nodata & \nodata & \nodata & $100\pm4$ & $30\pm3$ & $16\pm4$ & $13\pm4$ \\
X008 & $60\pm10$ & \nodata & \nodata & \nodata & \nodata & $100\pm6$ & $41\pm5$ & \nodata & \nodata \\
X009 & \nodata & \nodata & \nodata & \nodata & \nodata & $100\pm10$ & \nodata & \nodata & \nodata \\
X010 & \nodata & \nodata & \nodata & \nodata & \nodata & $100\pm10$ & \nodata & \nodata & \nodata \\
\nodata & & & & & & & & & \\
\enddata
\tablecomments{This table is organized into sections similarly to Table \ref{tab:obj}.  Extinction correction has not been applied. Line fluxes are reported relative to H$\beta$=100; however, line flux ratios for lines with large wavelength separations are unreliable for spectra where the ADC was not functioning (see Table \ref{tab:obj}).  For objects where H$\beta$ is not detected, fluxes are instead normalized relative to another line whose flux is given as 100.  Table \ref{tab:mea} is published in its entirety in the electronic edition of the {\it Astrophysical Journal} and at \url{https://www.cfa.harvard.edu/~nsanders/papers/M31/summary.html}.  A portion is shown here for guidance regarding its form and content.}
\end{deluxetable}
\clearpage
\begin{deluxetable}{lp{40pt}p{40pt}p{40pt}p{40pt}p{40pt}p{40pt}p{40pt}p{40pt}}
\tabletypesize{\scriptsize}
\tablecaption{Derived quantities for M31 HII regions \label{tab:derHII}}
\tablehead{\colhead{ID} & \colhead{$A_V$} & \colhead{Direct} & \colhead{Z94} & \colhead{KD02} & \colhead{N06 N2} & \colhead{N06 O3N2} & \colhead{PT05} & \colhead{PVT ONS}\\
 & \colhead{(mag)} & \colhead{(log(O/H)+12)} & \colhead{(log(O/H)+12)} & \colhead{(log(O/H)+12)} & \colhead{(log(O/H)+12)} & \colhead{(log(O/H)+12)} & \colhead{(log(O/H)+12)} & \colhead{(log(N/H)+12)}}
\startdata 
HII001 & \nodata & \nodata & \nodata & \nodata & $9.06\pm0.03$ & \nodata & \nodata & \nodata \\
HII002 & \nodata & \nodata & \nodata & \nodata & \nodata & \nodata & \nodata & \nodata \\
HII003 & \nodata & \nodata & \nodata & \nodata & $8.90\pm0.08$ & \nodata & \nodata & \nodata \\
HII004 & \nodata & \nodata & \nodata & \nodata & $8.87\pm0.09$ & \nodata & \nodata & \nodata \\
HII005 & \nodata & \nodata & \nodata & \nodata & \nodata & \nodata & \nodata & \nodata \\
HII006 & $0.6\pm0.8$ & \nodata & \nodata & \nodata & \nodata & \nodata & \nodata & \nodata \\
HII007 & \nodata & \nodata & \nodata & \nodata & \nodata & \nodata & \nodata & \nodata \\
HII008 & \nodata & \nodata & \nodata & \nodata & \nodata & \nodata & \nodata & \nodata \\
HII009 & $2.6\pm0.4$ & \nodata & \nodata & \nodata & \nodata & \nodata & \nodata & \nodata \\
HII010 & $0.2\pm0.3$ & \nodata & \nodata & $8.85\pm0.05$ & $8.83\pm0.06$ & \nodata & \nodata & \nodata \\
HII011 & $1.1\pm0.4$ & \nodata & \nodata & $8.76\pm0.08$ & $8.89\pm0.09$ & \nodata & \nodata & \nodata \\
HII012 & $0.6\pm0.5$ & \nodata & \nodata & $8.68\pm0.09$ & $8.70\pm0.07$ & $8.86\pm0.05$ & \nodata & \nodata \\
HII013 & $0.49\pm0.07$ & \nodata & $8.72\pm0.02$ & $8.74\pm0.01$ & $8.63\pm0.01$ & 8.63 & $8.23\pm0.02$ & $7.36\pm0.02$ \\
HII014 & $0.7\pm0.8$ & \nodata & \nodata & $8.8\pm0.1$ & $9.0\pm0.2$ & \nodata & \nodata & \nodata \\
HII015 & $0.6\pm0.4$ & \nodata & \nodata & $8.87\pm0.07$ & $9.01\pm0.09$ & \nodata & \nodata & \nodata \\
HII016 & \nodata & \nodata & \nodata & \nodata & \nodata & \nodata & \nodata & \nodata \\
HII017 & $0.5\pm0.6$ & \nodata & \nodata & \nodata & \nodata & \nodata & \nodata & \nodata \\
HII018 & \nodata & \nodata & \nodata & \nodata & \nodata & \nodata & \nodata & \nodata \\
HII019 & \nodata & \nodata & \nodata & \nodata & $9.0\pm0.1$ & \nodata & \nodata & \nodata \\
HII020 & \nodata & \nodata & \nodata & \nodata & $8.35\pm0.01$ & \nodata & \nodata & \nodata \\
HII021 & $1\pm1$ & \nodata & \nodata & $8.9\pm0.1$ & $8.9\pm0.2$ & \nodata & \nodata & \nodata \\
HII022 & \nodata & \nodata & \nodata & \nodata & \nodata & \nodata & \nodata & \nodata \\
HII023 & \nodata & \nodata & \nodata & \nodata & $9.03\pm0.01$ & \nodata & \nodata & \nodata \\
HII024 & \nodata & \nodata & \nodata & \nodata & $9.05\pm0.03$ & \nodata & \nodata & \nodata \\
HII025 & $2.1\pm0.6$ & \nodata & \nodata & \nodata & \nodata & \nodata & \nodata & \nodata \\
HII026 & $1.4\pm0.8$ & \nodata & \nodata & $8.8\pm0.1$ & $8.99\pm0.10$ & \nodata & \nodata & \nodata \\
HII027 & $0.7\pm0.4$ & \nodata & \nodata & $8.76\pm0.08$ & $8.85\pm0.09$ & \nodata & \nodata & \nodata \\
HII028 & $1.1\pm0.5$ & \nodata & \nodata & $8.81\pm0.09$ & $8.79\pm0.09$ & $8.93\pm0.07$ & \nodata & \nodata \\
HII029 & $0.97\pm0.09$ & \nodata & $8.94\pm0.02$ & $8.85\pm0.02$ & $8.78\pm0.02$ & $8.83\pm0.01$ & $8.29\pm0.04$ & $7.46\pm0.03$ \\
HII030 & $1.1\pm0.5$ & \nodata & \nodata & \nodata & $9.1\pm0.1$ & $8.77\pm0.05$ & \nodata & \nodata \\
HII031 & $2\pm1$ & \nodata & \nodata & \nodata & \nodata & \nodata & \nodata & \nodata \\
HII032 & $0.4\pm0.4$ & \nodata & \nodata & $8.82\pm0.06$ & \nodata & \nodata & \nodata & \nodata \\
HII033 & $0.6\pm0.5$ & \nodata & \nodata & $8.82\pm0.07$ & $9.1\pm0.1$ & \nodata & \nodata & \nodata \\
HII034 & $2.8\pm0.8$ & \nodata & \nodata & $8.6\pm0.2$ & $8.91\pm0.07$ & \nodata & \nodata & \nodata \\
HII035 & $0.1\pm0.2$ & \nodata & \nodata & \nodata & \nodata & \nodata & \nodata & \nodata \\
HII036 & $2.2\pm0.7$ & \nodata & \nodata & \nodata & $8.73\pm0.09$ & \nodata & \nodata & \nodata \\
HII037 & $1.0\pm0.7$ & \nodata & \nodata & $8.9\pm0.1$ & $9.1\pm0.1$ & \nodata & \nodata & \nodata \\
HII038 & $1.4\pm1.0$ & \nodata & \nodata & $8.9\pm0.1$ & $9.2\pm0.2$ & \nodata & \nodata & \nodata \\
HII039 & $0.7\pm0.5$ & \nodata & \nodata & \nodata & $8.82\pm0.09$ & $8.85\pm0.07$ & \nodata & \nodata \\
HII040 & $1.87\pm0.05$ & \nodata & 8.86 & 8.84 & $8.85\pm0.01$ & 8.76 & $8.30\pm0.01$ & $7.47\pm0.01$ \\
HII041 & $3.1\pm0.7$ & \nodata & \nodata & $8.8\pm0.1$ & $8.99\pm0.07$ & \nodata & \nodata & \nodata \\
HII042 & $2.0\pm0.4$ & \nodata & \nodata & \nodata & $8.8\pm0.1$ & $8.88\pm0.05$ & \nodata & \nodata \\
HII043 & $1.5\pm0.9$ & \nodata & \nodata & $8.9\pm0.1$ & \nodata & \nodata & \nodata & \nodata \\
HII044 & \nodata & \nodata & \nodata & \nodata & \nodata & \nodata & \nodata & \nodata \\
HII045 & $0.59\pm0.05$ & $8.33\pm0.04$ & $8.68\pm0.01$ & 8.77 & 8.40 & 8.39 & 8.46 & $7.39\pm0.02$ \\
HII046 & $0.8\pm0.2$ & \nodata & $8.96\pm0.04$ & $8.88\pm0.03$ & $8.86\pm0.02$ & $8.86\pm0.01$ & $8.32\pm0.06$ & $7.53\pm0.05$ \\
HII047 & 0.0 & \nodata & \nodata & $9.18\pm0.01$ & \nodata & \nodata & \nodata & \nodata \\
HII048 & $0.33\pm0.04$ & \nodata & 8.96 & 8.87 & 8.55 & 8.62 & 8.56 & $7.57\pm0.01$ \\
HII049 & $1.2\pm1.0$ & \nodata & \nodata & $8.7\pm0.2$ & $8.8\pm0.2$ & \nodata & \nodata & \nodata \\
HII050 & $3.0\pm0.6$ & \nodata & \nodata & \nodata & $9.10\pm0.07$ & \nodata & \nodata & \nodata \\
\nodata & & & & & & & & \\
\enddata
\tablecomments{The abundance diagnostics applied in this table are described in \S \ref{sec:obs-abundance}.  The reported uncertainties are derived by propagation of the line flux uncertainties, as described in \S \ref{sec:obs-abundance} and does not include other systematic effects.  When this ``statistical'' uncertainty is less than 0.01 dex, we do not report it.    Table \ref{tab:derHII} is published in its entirety in the electronic edition of the {\it Astrophysical Journal} and at \url{https://www.cfa.harvard.edu/~nsanders/papers/M31/summary.html}.  A portion is shown here for guidance regarding its form and content.}
\end{deluxetable}
\clearpage
\begin{deluxetable}{lp{40pt}p{40pt}}
\tabletypesize{\scriptsize}
\tablecaption{Derived quantities for M31 planetary nebulae \label{tab:derPN}}
\tablehead{\colhead{ID} & \colhead{$A_V$} & \colhead{log(O/H)+12}}
\startdata 
PN001& \nodata& \nodata \\
PN002& $1.3\pm0.1$& \nodata \\
PN003& $0.3\pm0.2$& \nodata \\
PN004& $0.09\pm0.05$& $8.51\pm0.04$ \\
PN005& $0.4\pm0.5$& \nodata \\
PN006& $1.5\pm0.3$& \nodata \\
PN007& $0.4\pm0.2$& \nodata \\
PN008& $0.6\pm0.3$& \nodata \\
PN009& \nodata& \nodata \\
PN010& \nodata& \nodata \\
PN011& \nodata& \nodata \\
PN012& $0.7\pm0.5$& \nodata \\
PN013& 0.0& \nodata \\
PN014& $0.01\pm0.09$& \nodata \\
PN015& $1.8\pm0.3$& \nodata \\
PN016& $0.5\pm0.7$& \nodata \\
PN017& $0.4\pm0.4$& \nodata \\
PN018& $0.0\pm0.4$& \nodata \\
PN019& $0.6\pm0.2$& \nodata \\
PN020& $2.0\pm0.1$& $7.6\pm0.2$ \\
PN021& $0.1\pm0.1$& \nodata \\
PN022& $1.0\pm0.4$& \nodata \\
PN023& $2\pm1$& \nodata \\
PN024& $1.1\pm0.9$& \nodata \\
PN025& \nodata& \nodata \\
\nodata & & \\
PNh001& $0.1\pm0.1$& \nodata \\
PNh002& \nodata& \nodata \\
PNh003& $0.2\pm0.4$& \nodata \\
PNh004& 0.0& \nodata \\
PNh005& \nodata& \nodata \\
PNh006& $0.1\pm0.2$& \nodata \\
PNh007& \nodata& \nodata \\
PNh008& 0.0& $8.7\pm0.1$ \\
PNh009& \nodata& \nodata \\
PNh010& \nodata& \nodata \\
PNh011& 0.0& \nodata \\
PNh012& $0.17\pm0.07$& \nodata \\
PNh013& 0.0& \nodata \\
PNh014& \nodata& \nodata \\
PNh015& $0.3\pm0.4$& \nodata \\
PNh016& $0.3\pm0.3$& \nodata \\
PNh017& $0.01\pm0.03$& \nodata \\
PNh018& $0.2\pm0.1$& \nodata \\
PNh019& $0.02\pm0.06$& \nodata \\
PNh020& $0.22\pm0.10$& $8.61\pm0.07$ \\
PNh021& \nodata& $8.62\pm0.05$ \\
PNh022& 0.0& \nodata \\
PNh023& \nodata& \nodata \\
PNh024& \nodata& \nodata \\
PNh025& $0.1\pm0.2$& \nodata \\
\nodata & &\\
\enddata
\tablecomments{See caption for Table~\ref{tab:derHII}.}
\end{deluxetable}
\clearpage
\begin{deluxetable}{lccccccccc}
\tabletypesize{\scriptsize}
\tablecaption{Fitted parameters for the abundance profiles of M31 HII regions and PNe\label{tab:fitparam}}
\tablehead{ & N & \multicolumn{2}{c}{Bootstrap} & \multicolumn{2}{c}{Spearman} \\
 &  & \colhead{Central} & \colhead{Slope} & $\rho$ & p \\
 &  & (dex) & (dex kpc$^{-1})$ &  & }
\startdata 
\cutinhead{HII regions --- All}
$A_V$ & 199 & $2.31 \pm 0.26 $  & $-0.0690 \pm 0.0168 $  & -0.31 & $8\times10^{-06}$ \\
log([N II]/H$\alpha$) & 223 & $-0.42 \pm 0.06 $  & $-0.0087 \pm 0.0048 $  & -0.17 & $0.01$ \\
$R_{23}$ & 61 & $0.43 \pm 0.06 $  & $0.0169 \pm 0.0044 $  & 0.54 & $8\times10^{-06}$ \\
$P$ & 61 & $0.26 \pm 0.08 $  & $0.0077 \pm 0.0061 $  & 0.18 & $0.18$ \\
log(O/H)+12 (Z94) & 60 & $9.10 \pm 0.06 $  & $-0.0208 \pm 0.0048 $  & -0.57 & $2\times10^{-06}$ \\
log(O/H)+12 (KD02) & 136 & $8.96 \pm 0.06 $  & $-0.0096 \pm 0.0049 $  & -0.33 & $1\times10^{-04}$ \\
log(O/H)+12 (N06 N2) & 192 & $9.13 \pm 0.07 $  & $-0.0195 \pm 0.0055 $  & -0.26 & $3\times10^{-04}$ \\
log(O/H)+12 (N06 O3N2) & 100 & $8.98 \pm 0.08 $  & $-0.0130 \pm 0.0068 $  & -0.19 & $0.06$ \\
log(O/H)+12 (PT05) & 48 & $8.42 \pm 0.09 $  & $-0.0054 \pm 0.0064 $  & -0.08 & $0.60$ \\
log(N/H)+12 (PVT ONS) & 52 & $7.83 \pm 0.07 $  & $-0.0303 \pm 0.0049 $  & -0.53 & $5\times10^{-05}$ \\
\cutinhead{HII regions --- Stellar}
$A_V$ & 92 & $2.20 \pm 0.36 $  & $-0.0555 \pm 0.0231 $  & -0.27 & $0.01$ \\
log([N II]/H$\alpha$) & 98 & $-0.34 \pm 0.07 $  & $-0.0080 \pm 0.0056 $  & -0.14 & $0.18$ \\
$R_{23}$ & 14 & $0.54 \pm 0.12 $  & $0.0107 \pm 0.0096 $  & 0.52 & $0.06$ \\
$P$ & 14 & $0.04 \pm 0.11 $  & $0.0218 \pm 0.0068 $  & 0.55 & $0.04$ \\
log(O/H)+12 (Z94) & 14 & $9.00 \pm 0.16 $  & $-0.0152 \pm 0.0133 $  & -0.52 & $0.06$ \\
log(O/H)+12 (KD02) & 47 & $8.92 \pm 0.12 $  & $-0.0037 \pm 0.0092 $  & -0.16 & $0.27$ \\
log(O/H)+12 (N06 N2) & 72 & $9.24 \pm 0.11 $  & $-0.0224 \pm 0.0082 $  & -0.39 & $6\times10^{-04}$ \\
log(O/H)+12 (N06 O3N2) & 25 & $9.10 \pm 0.13 $  & $-0.0191 \pm 0.0110 $  & -0.33 & $0.10$ \\
log(O/H)+12 (PT05) & 10 & $8.18 \pm 0.19 $  & $0.0077 \pm 0.0141 $  & 0.33 & $0.35$ \\
log(N/H)+12 (PVT ONS) & 10 & $7.80 \pm 0.16 $  & $-0.0283 \pm 0.0112 $  & -0.42 & $0.23$ \\
\cutinhead{HII regions --- Diffuse}
$A_V$ & 107 & $2.75 \pm 0.37 $  & $-0.1066 \pm 0.0261 $  & -0.40 & $2\times10^{-05}$ \\
log([N II]/H$\alpha$) & 125 & $-0.37 \pm 0.06 $  & $-0.0177 \pm 0.0049 $  & -0.26 & $4\times10^{-03}$ \\
$R_{23}$ & 47 & $0.38 \pm 0.07 $  & $0.0210 \pm 0.0044 $  & 0.52 & $2\times10^{-04}$ \\
$P$ & 47 & $0.33 \pm 0.11 $  & $0.0033 \pm 0.0089 $  & 0.08 & $0.60$ \\
log(O/H)+12 (Z94) & 46 & $9.16 \pm 0.07 $  & $-0.0248 \pm 0.0048 $  & -0.56 & $6\times10^{-05}$ \\
log(O/H)+12 (KD02) & 89 & $9.01 \pm 0.04 $  & $-0.0160 \pm 0.0031 $  & -0.43 & $2\times10^{-05}$ \\
log(O/H)+12 (N06 N2) & 120 & $9.09 \pm 0.09 $  & $-0.0195 \pm 0.0070 $  & -0.20 & $0.03$ \\
log(O/H)+12 (N06 O3N2) & 75 & $8.93 \pm 0.11 $  & $-0.0102 \pm 0.0095 $  & -0.11 & $0.35$ \\
log(O/H)+12 (PT05) & 38 & $8.51 \pm 0.09 $  & $-0.0113 \pm 0.0065 $  & -0.16 & $0.33$ \\
log(N/H)+12 (PVT ONS) & 42 & $7.83 \pm 0.08 $  & $-0.0300 \pm 0.0058 $  & -0.54 & $2\times10^{-04}$ \\
\cutinhead{PNe --- Disk}
$A_V$ & 333 & $0.58 \pm 0.08 $  & $0.0070 \pm 0.0071 $  & 0.04 & $0.51$ \\
log([N II]/H$\alpha$) & 277 & $-0.57 \pm 0.05 $  & $-0.0023 \pm 0.0044 $  & -0.05 & $0.38$ \\
$R_{23}$ & 148 & $1.20 \pm 0.03 $  & $-0.0016 \pm 0.0027 $  & -0.07 & $0.42$ \\
$P$ & 148 & $0.93 \pm 0.02 $  & $-0.0027 \pm 0.0020 $  & -0.17 & $0.04$ \\
log(O/H)+12 (direct) & 51 & $8.47 \pm 0.09 $  & $-0.0056 \pm 0.0076 $  & -0.11 & $0.45$ \\
\enddata
\tablecomments{The bootstrap fitting reported in this table is described in \S \ref{sec:disc-gradient}.  The central value and slope define the best-fit line to the data with respect to the de-projected radius in M31.}
\end{deluxetable}

\end{document}